\documentclass[reprint,amsfonts, amssymb, amsmath,  showkeys,aps,pra, superscriptaddress,twocolumn ,longbibliography,nofootinbib]{revtex4-2}

\usepackage{xcolor}

\usepackage{float}
\makeatletter
\let\newfloat\newfloat@ltx
\makeatother
\usepackage[english]{babel}
\usepackage[utf8]{inputenc}
\usepackage{graphics}
\usepackage{selinput}
\usepackage[normalem]{ulem}
\usepackage[shortlabels]{enumitem}

\usepackage{braket}
\usepackage{amsthm}
\usepackage{mathtools}
\usepackage{physics}
\usepackage{graphicx}
\usepackage[left=16mm,right=16mm,top=16mm,bottom=16mm,columnsep=15pt]{geometry} 
\usepackage{adjustbox}
\usepackage{placeins}
\usepackage[T1]{fontenc}
\usepackage{lipsum}
\usepackage{csquotes}
\usepackage{bm}

\usepackage{quantikz}

\usepackage[makeroom]{cancel}
\usepackage[toc,page]{appendix}
\usepackage[colorlinks=true,citecolor=blue,linkcolor=magenta]{hyperref}

\usepackage{algorithm}
\usepackage{algpseudocode}

\usepackage{tikz}
\tikzset{every picture/.style=remember picture}

\usepackage[utf8]{inputenc}
\usepackage{graphicx}
\usepackage{xcolor}
\usepackage{amsmath}
\usepackage{amsthm}
\usepackage{bm}
\usepackage{booktabs}
\usepackage{bbm}
\usepackage{comment}
\usepackage{appendix}
\usepackage{mathdots}
\usepackage{lipsum}
\usepackage{verbatim}
\usepackage{natbib}
\usepackage{nccmath}
\usepackage{multirow}
\usepackage{amsfonts} 

\newtheorem{action}{Action}
\newtheorem{problem}{Problem}
\newtheorem{definition}{Definition}

\usepackage{xspace}
\newcommand{\methodname}{{\textsc{Q-PreSyn}}\xspace}
\newcommand{\rrangle}{\rangle\!\rangle}
\newcommand{\llangle}{\langle\!\langle}

\makeatletter

\DeclareRobustCommand\bra[1]{%
  \@ifnextchar\ket{\br@k@t{#1}}{\br@{#1}}%
}
\newcommand\br@[1]{{\langle{#1}\lvert}}

\DeclareRobustCommand\ket[1]{%
  \@ifnextchar\bra{\k@t{#1}\!}{\k@t{#1}}%
}
\newcommand\k@t[1]{{\lvert{#1}\rangle}}

\newcommand\br@k@t[1]{{\langle{#1}}}

\DeclareRobustCommand\bbra[1]{%
  \@ifnextchar\kket{\bbr@kk@t{#1}}{\bbr@{#1}}%
}
\newcommand\bbr@[1]{{\llangle{#1}\lvert}}

\DeclareRobustCommand\kket[1]{%
  \@ifnextchar\bbra{\kk@t{#1}\!}{\kk@t{#1}}%
}
\newcommand\kk@t[1]{{\lvert{#1}\rrangle}}

\newcommand\bbr@kk@t[1]{{\llangle{#1}}}

\usepackage{amssymb}
\usepackage{dsfont}
\usepackage{csvsimple}

\def\be{\begin{equation}}
\def\te{\end{equation}}
\def\ee{\end{equation}}
\def\ba{\begin{eqnarray}}
\def\bea{\begin{eqnarray}}

\def\tea{\end{eqnarray}}
\def\ea{\end{eqnarray}}
\def\eea{\end{eqnarray}}

\newcommand{\calc}{\mathcal{C}}
\newcommand{\cala}{\mathcal{A}}
\newcommand{\call}{\mathcal{L}}

\begin{document}

\title{Quantum Circuit Pre-Synthesis: Learning Local Edits to Reduce $T$-count}

\author{Daniele {Lizzio Bosco}}
\affiliation{Department of Mathematics, Computer Science and Physics, University of Udine, Udine, Italy}
\affiliation{Department of Biology, University of Naples Federico II, Naples, Italy}

\author{Lukasz Cincio}
\affiliation{Theoretical Division, Los Alamos National Laboratory, Los Alamos, New Mexico 87545, USA}
\affiliation{Quantum Science Center, Oak Ridge, TN 37931, USA}

\author{Giuseppe Serra}
\affiliation{Department of Mathematics, Computer Science and Physics, University of Udine, Udine, Italy}

\author{M. Cerezo}
\thanks{cerezo@lanl.gov}
\affiliation{Information Sciences, Los Alamos National Laboratory, Los Alamos, New Mexico 87545, USA}
\affiliation{Quantum Science Center, Oak Ridge, TN 37931, USA}

\begin{abstract}

Compiling quantum circuits into Clifford+$T$ gates is a central task for fault-tolerant quantum computing using stabilizer codes. In the near term, $T$ gates will dominate the cost of fault tolerant implementations, and any reduction in the number of such expensive gates could mean the difference between being able to run a circuit or not. While exact synthesis is exponentially hard in the number of qubits, local synthesis approaches are commonly used to compile large circuits by decomposing them into substructures. However, composing local methods leads to suboptimal compilations in key metrics such as $T$-count or circuit depth, and their performance strongly depends on circuit representation. In this work, we address this challenge by proposing \textsc{Q-PreSyn}, a strategy that, given a set of local edits preserving circuit equivalence, uses a RL agent to identify effective sequences of such actions and thereby obtain circuit representations that yield a reduced $T$-count upon synthesis. Experimental results of our proposed strategy, applied on top of well-known synthesis algorithms, show up to a $20\%$ reduction in $T$-count on circuits with up to 25 qubits, without introducing any additional approximation error prior to synthesis.
\end{abstract}
\maketitle

\section{Introduction}
\label{sec:intro}

\begin{figure*}[hbt]
    \centering
    \includegraphics[width=1\linewidth]{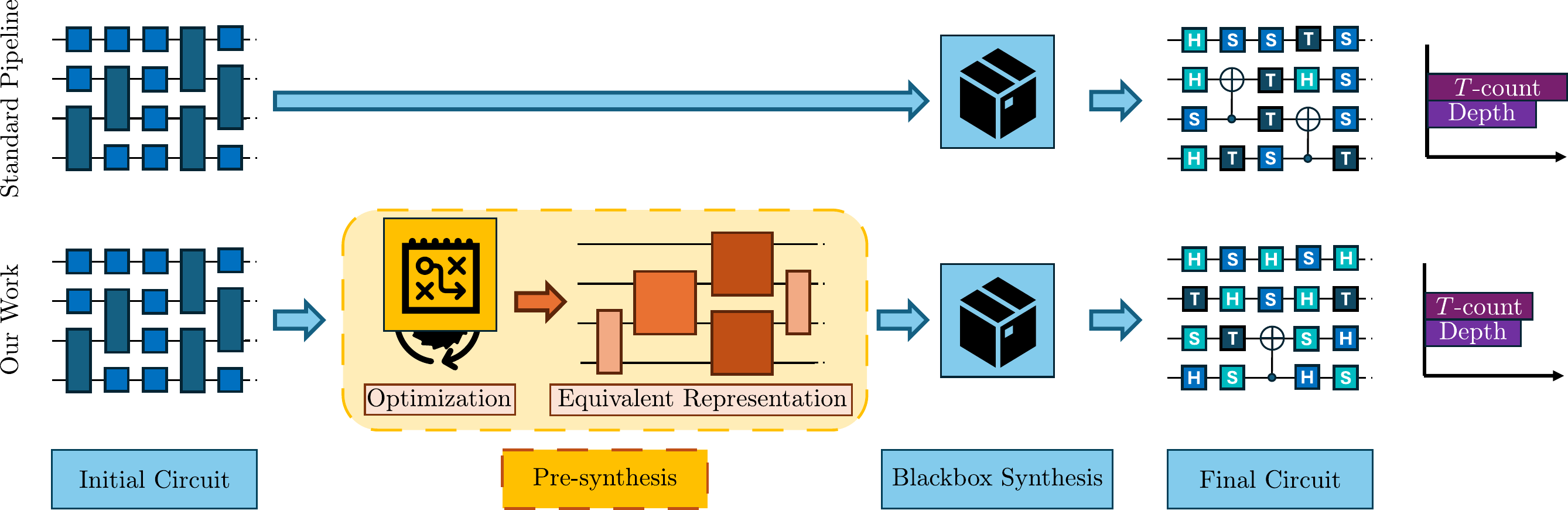}
    \caption{\textbf{Description of \methodname applied in a default synthesis pipeline.} Starting from the initial circuit representation, given a blackbox local synthesis algorithm, we explore the space of equivalent representations to find the ones that lead to a reduced $T$-count after the synthesis. We show that our methodology reduces both the number of $T$ }
    \label{fig:visual-teaser}
\end{figure*}

Fault-tolerant quantum computation relies on the ability to efficiently decompose  unitary operations of interest into sequences of gates drawn from a discrete set such as the Clifford+$T$ gates. This process, known as \emph{quantum circuit synthesis}~\cite{kitaev_quantum_1997, amy_meet_middle_2013, kliuchnikov_fast_2013, davis_towards_2020, he_unitary_2023}, bridges the gap between high-level algorithmic descriptions and low-level hardware implementations. 
The direct decomposition of a target $n$-qubit unitary into elementary gates, usually called global synthesis, is computationally intractable for all but the smallest systems, with worst-case complexity scaling exponentially in the number of qubits~\cite{quantum_synthesis_quer_2021, sun_asymptotically_2023}. 
As a result, practical compilers employ \emph{local} strategies, decomposing large circuits into smaller subcircuits that are synthesized independently and later recombined~\cite{ datastructure, Bravyi2021cliffordcircuit,gouzien2025provably}.

While local synthesis dramatically improves scalability, it introduces a key limitation: the quality of the resulting decomposition often depends sensitively on the \emph{representation} of the circuit (i.e. its factorization into smaller gates) before the synthesis process. A pathological illustration of this phenomenon arises in the case of consecutive single-qubit rotations. Consider the task of synthesizing a single-qubit $R_z(\alpha)$ rotation in Clifford+$T$ gates. Given an approximation tolerance $\varepsilon$, this compilation can be performed near-optimally, in the number of $T$-gates, using algorithms such as  \texttt{gridsynth}~\cite{gridsynth}. 
However, if the circuit representation contains two consecutive $R_z(\alpha_1)$ and $R_z(\alpha_2)$ gates and one synthesize them independently, the resulting decomposition has typically double of the number of required $T$ gates relative to synthesizing their combination $R_z(\alpha_1+\alpha_2)$ directly. This na\"ive examples shows that equivalent circuit representations can yield vastly different $T$-counts (i.e., number of $T$-gates), a central metric in fault-tolerant synthesis~\cite{magic_state_t}.  Similarly, equivalent representations obtained by merging or regrouping multi-qubit gates can lead to substantially more efficient synthesized circuits. This observation motivates the following question:
\emph{Given a quantum circuit and a local synthesis algorithm, can we determine a circuit representation that enables more efficient synthesis, leading to lower $T$-counts?}

To address this question, we explore the space of equivalent circuit representations through merge operations that preserve the overall unitary implemented by the circuit but modify its local structure, thereby influencing the outcome of subsequent synthesis.
To efficiently explore this space, we formulate the task of minimizing the final $T$-count as a planning problem, and we employ reinforcement learning~\cite{intro_RL} (RL) to discover strategies that yield lower synthesis costs (see Fig.~\ref{fig:visual-teaser}). 
Our heuristic results highlight the potential of learning-based optimization for quantum compilation, showing consistent improvements in post-synthesis efficiency across  Clifford+$T$ synthesis of general unitaries, real-time evolutions~\cite{kukliansky2024leveraging,zhang2024scalable,gibbs2025learning}, matchgate synthesis~\cite{matchgate_circuit}, and  throughout different circuit structures and number of qubits. 

Overall, this work contributes a new perspective on quantum synthesis: It treats circuit representation as an optimizable object, distinct from synthesis itself, through our proposed method \methodname.
\methodname shows that learning-guided structural transformations can yield significant gains in synthesis efficiency. Our proposed method can be applied on top of any synthesis algorithm, and can thus serve as a universal preprocessing, or \textit{pre-synthesis}, stage compatible with a wide range of compilation pipelines, including \texttt{gridsynth}~\cite{gridsynth} and the Solovay-Kitaev~\cite{SK_alg} algorithm.

In summary, the main contributions of this work are:
\begin{enumerate}
    \item We present \methodname, an approach for \emph{representation-level optimization} of quantum circuits, in which circuit structure is modified through unitary-preserving merge operations to improve the quality of later syntheses.
    \item We formalize the problem of identifying advantageous merge sequences as a \emph{plan optimization task}, where each plan corresponds to a sequence of local structural edits. A RL-based optimization strategy can then be used to explore the plan space by receiving feedback from synthesis outcomes, outperforming greedy strategies and suggesting long-term dependencies between merges. 
    \item We evaluate the proposed pipeline across different synthesis frameworks, combining {Qiskit}~\cite{qiskit} and \texttt{BQSKit}~\cite{BQSKit} with well-known algorithms such as \texttt{gridsynth}~\cite{gridsynth} and Solovay-Kitaev~\cite{SK_alg}, achieving up to a $20\%$ reduction in $T$-count in most scenarios.
    \item To facilitate further research, we provide all code necessary to reproduce our method and evaluate the experiments~\cite{lizzio2025quantum}.
\end{enumerate}

The remainder of this paper is organized as follows. Section~\ref{sec:RW} reviews prior work in circuit optimization and reinforcement-learning–based compilation. In Section~\ref{sec:approx_synthesis} we present the background for local approximate synthesis. 
  Section~\ref{sec:method} introduces the proposed framework. Section ~\ref{sec:exp} describes our experimental setup and implementation details. In Section ~\ref{sec:results} we present and discuss the results, and Section \ref{sec:conclusion} concludes with perspectives for future research.

\section{Results}

\subsection{Related Work}
\label{sec:RW}

In this work, we do not address directly the task of synthesis. Instead, we target the representation stage that precedes synthesis: Given a circuit and any off-the-shelf local compiler, we seek unitary-preserving structural edits that change how subcircuits are defined and therefore how they are subsequently combined by the compiler. Hence, our work is related to Quantum Circuit Optimization (QCO).

One of the first approaches for QCO is given by \textit{peephole} optimization, a technique that operates by inspecting small subcircuits (or \emph{peepholes}), synthesizing these sections independently, and replacing them with optimized equivalents \cite{ datastructure, Bravyi2021cliffordcircuit}. These approaches effectively reduce local redundancy and gate count by exploiting known equivalences in limited contexts.

More recent work extends these ideas to exact and approximate synthesis methods \cite{QGo, top-as, Quest, Quest2}. While exact synthesis aims to minimize cost metrics such as $T$-count or depth within a given gate set, approximate synthesis allows small errors in unitary representation to achieve more efficient circuits, often using stochastic or annealing-based search strategies.

Our work is related to these approaches at a high level, but differs in that we perform \emph{exact} structural modifications that preserve circuit equivalence without performing synthesis directly. Instead, we modify the circuit representation so that, when synthesis is later applied, it yields improved results.
The proposed framework treats the exploration of equivalent circuit representations as a RL task. This makes our approach \emph{synthesis-agnostic}, since it can be applied independently of the particular synthesis algorithm later employed.

More broadly, several works have introduced learning-based approaches to circuit optimization. In \cite{Quarl}, the authors employ graph neural networks to learn structural transformations improving Clifford+$T$ circuits. In \cite{fösel2021quantumcircuitoptimizationdeep}, the authors use RL to optimize parameterized quantum circuits directly. The approach proposed in \cite{RLxZXC} leverages ZX-calculus simplifications guided by RL to minimize Clifford+$T$ count. Finally, \cite{QCGame} formulates compilation as a multi-agent game, enabling general-purpose circuit reconfiguration. These efforts demonstrate the growing interest in RL-based meta-optimization strategies for quantum compilation and motivate our focus on learning structural transformations prior to synthesis.

Finally, despite not being directly related to our work, it is interesting to note that several RL-based strategies have been explored for the addressing directly local synthesis~\cite{RL_x_VQC_21, moro_quantum_2021, quetschlich_compiler_2023, rosenhahn_monte_2023, wang2024quantumcompilingreinforcementlearning, kremer2025optimizingnoncliffordcountunitarysynthesis, alphatensor_quantum,bilkis2021semi}, including a deep Q-network trained to sequentially select gates to approximate a given three-qubit unitary \cite{wang2024quantumcompilingreinforcementlearning},
 an agent that uses channel representations to synthesize two-qubit unitaries in the Clifford+$T$ basis \cite{kremer2025optimizingnoncliffordcountunitarysynthesis}, 
and a tensor-based RL method to discover exact synthesis procedures for Clifford+$T$ subsets admitting polynomial representations \cite{alphatensor_quantum}.

\subsection{Approximate Circuit Synthesis}
\label{sec:approx_synthesis}

Approximate circuit synthesis is the task of mapping a target unitary $U$ into a quantum circuit over a finite gate set $G$, such as $\{H,\,T,\,CNOT\}$ for near-term noisy computing, or Clifford+$T$ for fault-tolerant computation based on stabilizer codes\footnote{Mathematically, we pick a gate set $G$ that is dense in some group $\mathbb{G}\subseteq\mathcal{U}(2^n)$ from which $U$ belongs. For instance, $\{H,\,T,\,CNOT\}$, with CNOTs acting on nearest-neighbors, and Clifford+$T$ are dense in  $\mathcal{U}(2^n)$, allowing for universal synthesis. However, one can also pick gate sets that are dense in subgroups of the unitary group, such as when performing matchgate synthesis via matchgate-Clifford+$T$ synthesis~\cite{casas2025matchgate}. Here $\mathbb{G}$ is the $\mathbb{SPIN}$ representation of $\mathcal{SO}(2n)$. }. This problem can be formulated as 

\begin{problem}[Approximate Synthesis]
    Let $\calc$ be a quantum circuit implementing a unitary operation $U \in \mathcal{U}(2^n)$, and let $G$ be a gate set. The problem of \emph{approximate synthesis} consists of finding a circuit $\calc'$ expressed as a product of gates from $G$, such that $\calc'$ implements a unitary $U' \in \mathcal{U}(2^n)$ satisfying (potentially up to a global phase)
    \[
        d(U, U') \leq \varepsilon,
    \]
    where $d(\cdot, \cdot)$ denotes a suitable distance measure on unitaries, and $\varepsilon > 0$ is a given approximation tolerance. 
\end{problem}
\noindent Common choices of $d$ are the trace distance and the norm operator distance.

Among all circuits that synthesize a unitary $U$, we are often interested in finding the one that minimize a certain \textit{cost}. In noisy-quantum computation we could be interested in minimizing the number of number of CNOT gates or the depth of the circuit, as these reduce the detrimental effects of hardware errors. In fault tolerant computing we want to minimize the number of non-native gates, which require expensive injection or synthesis procedures to implement them within the code~\cite{bravyi2005universal,horsman2012surface,campbell2017roads}. In this work, we take $G$ to be Clifford$+T$ and we select the cost to be the number of $T$ gates, also called \textit{$T$-count}, which plays a central role in stabilizer code-based fault-tolerant devices.

In general, the cost of performing  global synthesis scales exponentially with the number of qubits $n$. For instance, when using the Solovay-Kitaev (SK) theorem~\cite{kitaev_quantum_1997} to compile a unitary from $\mathcal{U}(2^n)$ to $\varepsilon$ precision, one requires a number of gates  scaling as  $\Omega\left(2^n \log\left(\varepsilon^{-1}\right)/\log(n)\right)$ and $\mathcal{O}\left(n^2 4^n \log\left(n^2 4^n \varepsilon^{-1}\right)\right)$~\cite{nielsen2000quantum}. This means that the cost of synthesizing unitaries acting on large number of qubits quickly becomes intractable. This exponential scaling is avoided by instead first decomposing the target unitary $U$ into a product local unitaries acting on just a few qubits (typically acting on one or two qubits), synthesizing the local gates and then reconstructing $U$ by composition of the locally compiled unitaries.

At this point, we make several important remarks. First, we can see that if $U$ is expressed as the product of $\mathcal{O}({\rm poly}(n))$ local unitaries acting on $\mathcal{O}(1)$ qubits, then even if we perform the local synthesis with the SK theorem, the number of gates after the synthesis will only scale as  $\mathcal{O}({\rm poly}(n))$ (as per the results stated above). Hence, expressing $U$ in terms of unitaries avoids the exponential scaling induced by global compilation. Of course, this tradeoff is not free as there are two costs one pays when performing local versus global synthesis. The first arises from the fact that there are typically many representation for $U$ in terms of local gates (a fact exploited in this work), and since different synthesis generally lead to different gate counts, care must be taken in the representation choice. The second cost comes from the fact that if each local gate is compiled to an error   $\varepsilon_L$, then these errors add-up when reconstructing the global unitary. For instance, if $U$ is decomposed into $K\in\mathcal{O}({\rm poly}(n))$ local gates, then the total synthesis error can be as large as $K \varepsilon_L$ (via a simple triangular inequality). Thus, to ensure that the global synthesis is of the order $\varepsilon$, one must choose $\varepsilon_L\leq \varepsilon/K$. Luckily, even when using the SK theorem this only translates into a gate-depth increase of $\mathcal{O}(\log(n))$ gates per local synthesis. Taken together, one can see that despite the issue of the representation choice of $U$, there are many advantages for local versus global compilation.

\subsection{Pre-synthesis via reinforcement learning}
\label{sec:method}

We now describe the key steps of our proposed method \methodname for circuit pre-synthesis.  We consider a quantum circuit $\calc$, representing a global unitary $U$, which we assume can be expressed as 
\begin{equation}
    U=\prod_{l=1}^KU_l\label{eq:decomp}
\end{equation}
where each $U_l$ acts on at most two qubits. Note that this hypothesis is not restrictive, as a general unitary can be always be decomposed in smaller components acting on a lower number of qubits. Crucially, we remark that a decomposition of the form in Eq.~\eqref{eq:decomp} is precisely the output of state-of-the art modern techniques for compiling general unitaries $U$ into a product of local unitaries. Moreover, we note that standard techniques are based on some form of ansatz (tensor networks, parametrized quantum circuits, or other variable structure techniques~\cite{cartan_sur_1926, drury_constructive_2008, wierichs_recursive_2025,tang2019qubit,grimsley2019adaptive,cincio2018learning,moro2021quantum,bilkis2021semi,du2020quantum,sim2021adaptive,zhang2021mutual,tkachenko2020correlation,claudino2020benchmarking,rattew2019domain,chivilikhin2020mog,zhang2020differentiable,wada2022sequential,li2024quarl,fosel2021quantum,nakaji2025quantum,rosenhahn2023monte,kukliansky2024leveraging,zhang2024scalable}), leading to decompositions that are not optimal and can contain redundancies. For instance, a common strategy is to compile $U$ with tensor networks via matrix product operator techniques, where local tensors are individually optimized. This can lead to two adjacent unitaries $U_l$ and $U_l'$ acting on the same subset of qubits, but containing some redundancies. This is precisely the fact that we will exploit next. 

Then, as previously mentioned, our ultimate goal is to obtain an efficient decomposition of $\calc$ into gates from a given set $\mathcal{G}$ (e.g., Clifford+$T$), by leveraging a (blackbox) local synthesis algorithm $\mathcal{A}$, 
which approximates each local gate $U_l$ by $\mathcal{A}(U_l)$ within a chosen tolerance $\varepsilon$. In some cases, $\varepsilon$ can be treated as a tunable parameter of $\cala$. Moreover, the algorithm $\cala$ may also expose additional settings (for instance, a maximum runtime budget in \texttt{BQSKit} \cite{BQSKit}). Note also that $\cala$ can be non-deterministic. For the purposes of this discussion, we can assume that all the parameters of $\cala$ are fixed.

Putting it all together, from the decomposition $U=\prod_{l=1}^KU_l$, we can employ $\cala$ to obtain a global approximate compilation $\tilde\calc$ of $\calc$ as
\begin{equation}
    \cala(\calc)\coloneq \prod_{l=1}^K\cala(U_l). 
\end{equation}

\subsubsection{Actions}

In practice, different but equivalent representations of $U$ into products of local gates can yield decompositions with different properties (e.g., different depth or $T$-count). 
As a starting point, consider the task of synthesizing a single $R_z(\alpha)$ gate, for $\alpha\in[0,2\pi)$. 
For a fixed tolerance $\varepsilon$, the \textsc{Gridsynth}~\cite{gridsynth} algorithm produces a decomposition in Clifford+$T$ that is optimal in the number of $T$ gates. 
Now, consider a circuit composed of two adjacent $R_z$ rotations (e.g., $R_z(\alpha_1)$ and $R_z(\alpha_2)$). 
Since the $T$-count does not depend on the value of $\alpha$, applying \textsc{Gridsynth} separately to $R_z(\alpha_1)$ and $R_z(\alpha_2)$ yields, on average, twice the $T$-count compared to synthesizing $R_z(\alpha_1+\alpha_2)$ directly. 
In the extreme case $\alpha_1 = -\alpha_2$, the resulting unitary is equivalent to the identity, which requires no gates at all.
This motivates the following \textbf{merge} operation.
\begin{action}[\textsc{Single Qubit Merge $(i)$}]
    Let $\mathcal{C}$ be a circuit corresponding to a unitary $VU=\prod_{l=1}^K U_\ell$, and fix a qubit index $i$. 
    Consider a sequence of consecutive unitaries $U_{n_1},\dots, U_{n_k}$ acting exclusively on the qubit $i$.
    Replace them with the single gate 
    \[
    W = U_{n_k}\dots U_{n_1} \quad \in \mathcal{U}(2).
    \]
\end{action}
\noindent Note that, by construction, this action preserves the unitary implemented by the circuit. 

In analogy with the single-qubit case, we define an operation that merges consecutive two-qubit gates acting on a fixed pair of qubits, while also preserving the overall unitary implemented by the circuit.

\begin{figure}[t]
    \centering
    \includegraphics[width=1\columnwidth]{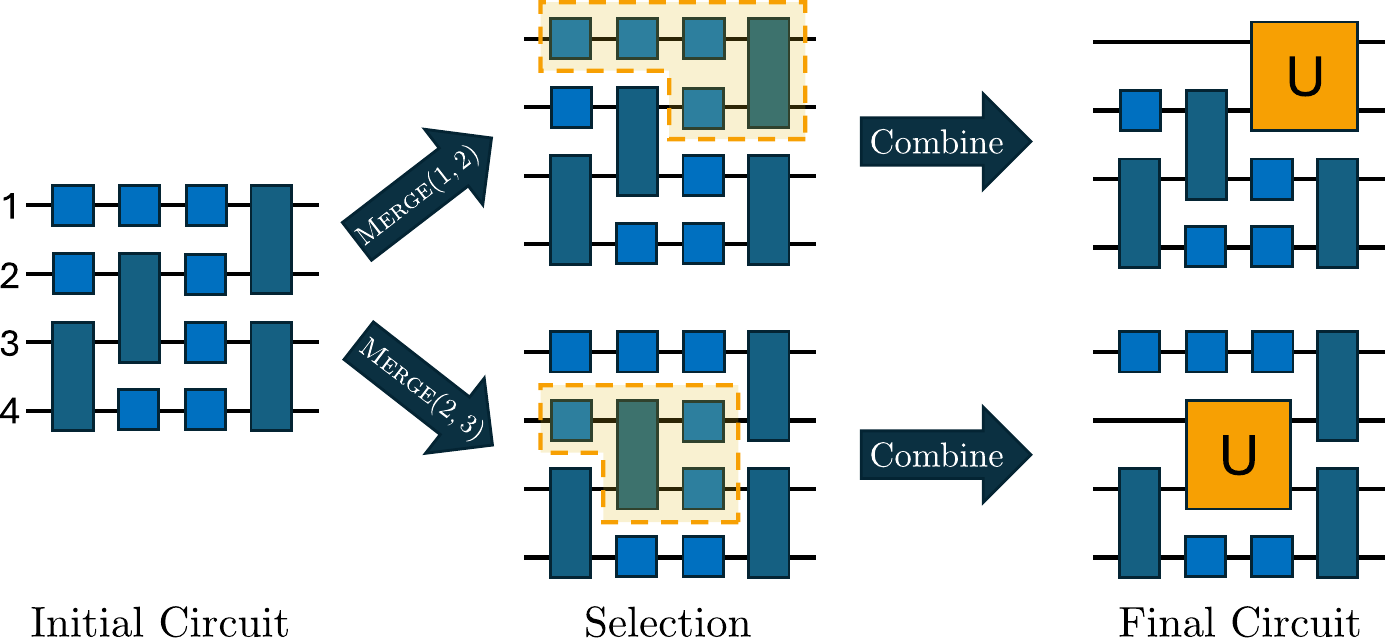}
    \caption{\textbf{Effect of the actions \textsc{Merge}$(1,2)$ and \textsc{Merge}$(2,3)$ on an example circuit}. First, the largest selection of gates is performed. Then, the selected gates are combined in a single unitary $U$. See Algorithm  \ref{alg:two_qubit_merge_contiguous} in the Methods Section for more details.}
    \label{fig:merge-scheme}
\end{figure}

\begin{action}[\textsc{Two Qubits Merge} $(i,j)$]
    Let $\mathcal{C}$ be a circuit corresponding to a unitary $U = \prod_{l=1}^K U_l$, and fix a pair of qubit indices $(i,j)$. 
    Consider a sequence of consecutive unitaries $U_{n_1}, \dots, U_{n_k}$ acting exclusively on the indices $i$, $j$, or both. 
    Replace them with the single gate
    \[
        W = U_{n_k} \dots U_{n_1} \quad \in \mathcal{U}(4).
    \]
\end{action}
A representation of the effect of this action is provided in Fig.~\ref{fig:merge-scheme} (also, see Algorithm \ref{alg:two_qubit_merge_contiguous} in the Methods Section).

In contrast to single qubit merges, the application of a two qubit merge is not always beneficial, and different sequences of merges can lead to vastly different results. As an example, consider the following pathological circuit:

\begin{center}
\begin{quantikz}
\lstick{$q_0$} & \qw & \gate[wires=2]{\,\,\,U\,\,\,} &\gate[wires=2]{U^{-1}} &  \qw & \qw & \qw \\
\lstick{$q_1$} & \gate{R_z(\alpha)}        &\qw     &                  & \gate{R_z(-\alpha)}              & \ctrl{1} & \qw \\
\lstick{$q_2$} & \qw              & \qw    &\qw           & \qw              & \targ{}    & \qw
\end{quantikz}
\end{center}

In this case, the action \textsc{Two Qubits Merge}$(0, 1)$ merges $R_z(\alpha)$, $U$, $U^{-1}$, and $R_z(-\alpha)$, obtaining a two-qubit identity. On the other hand, the action  \textsc{Two Qubit Merge}$(1, 2)$ combines together the gates $R_z(-\alpha)$ with the \textsc{CNOT} gate. After that, the merge on $(0, 1)$ does not simplify the  gates between $0$ and $1$ anymore, resulting in a more complex final representation. 

In general, it is possible to consider a more general \textsc{K~Qubits Merge} that acts on any number of qubits, by first obtaining the corresponding unitary $U$, and then synthesizing $U$ directly. This could be beneficial, as the synthesis of a larger subcircuit allows to find better decompositions than synthesizing smaller patches. However, the synthesis of circuits with more qubits is exponentially harder. Moreover, the space of possible merges of $k$ qubits between $n$ grows as $\binom{n}{k}$, making its exploration more complex. For these reasons, in this work we consider merges with up-to two qubits.

\subsubsection{Plans as sequences of actions}
The effect of applying a merge operation depends on the context in which it is performed. 
As illustrated in the example above, different orders of applying the same actions may lead to 
very different final representations of the circuit. 
To reason about this process systematically, we introduce the notion of a \emph{plan}.

\begin{definition}[Plan]
\label{def:plan}
    Let $\calc$ be an $n$-qubit circuit. 
    A \emph{plan} $\pi$ for $\calc$ is a finite sequence of actions
    \[
        \pi = (a_1, a_2, \ldots, a_m),
    \]
    where each $a_j$ is a \textsc{Two Qubit Merge}$(i,j)$ for some $i,j \in \{0,\dots,n-1\}$ with $i < j$. 
    Applying $\pi$ to $\calc$ yields a new equivalent representation $\calc_\pi$ obtained by executing the actions in order.
\end{definition}

 Given a local synthesis algorithm $\cala$ and a cost function $\call(\cdot)$ (e.g., the $T$-count after synthesis with $\cala$), 
the problem of \emph{plan selection} consists of finding a plan $\pi^*$ that minimizes the cost of the synthesized circuit:
\begin{equation}
\label{eq:formula_pistar}
    \pi^* \;=\; \arg\min_{\pi \in \Pi} \; \call\left(\cala\left(\calc_\pi\right)\right),
\end{equation}
where $\Pi$ denotes the space of all valid plans for $\calc$.  Note that when single qubit merges do not increase the cost $\call$ (e.g. the case with \texttt{gridsynth}  and $T$-count discussed above), we can always apply a single qubit merge for each qubit after the application of a plan $\pi$.

This formulation highlights two key challenges: 
(i) the combinatorial size of the plan space $\Pi$, 
and (ii) the fact that local merges can have non-local consequences, 
making greedy choices suboptimal. 
In practice, efficient heuristics or learned strategies are needed to explore this search space.

\begin{figure}[htbp]
    \centering
\includegraphics[width=1\columnwidth]{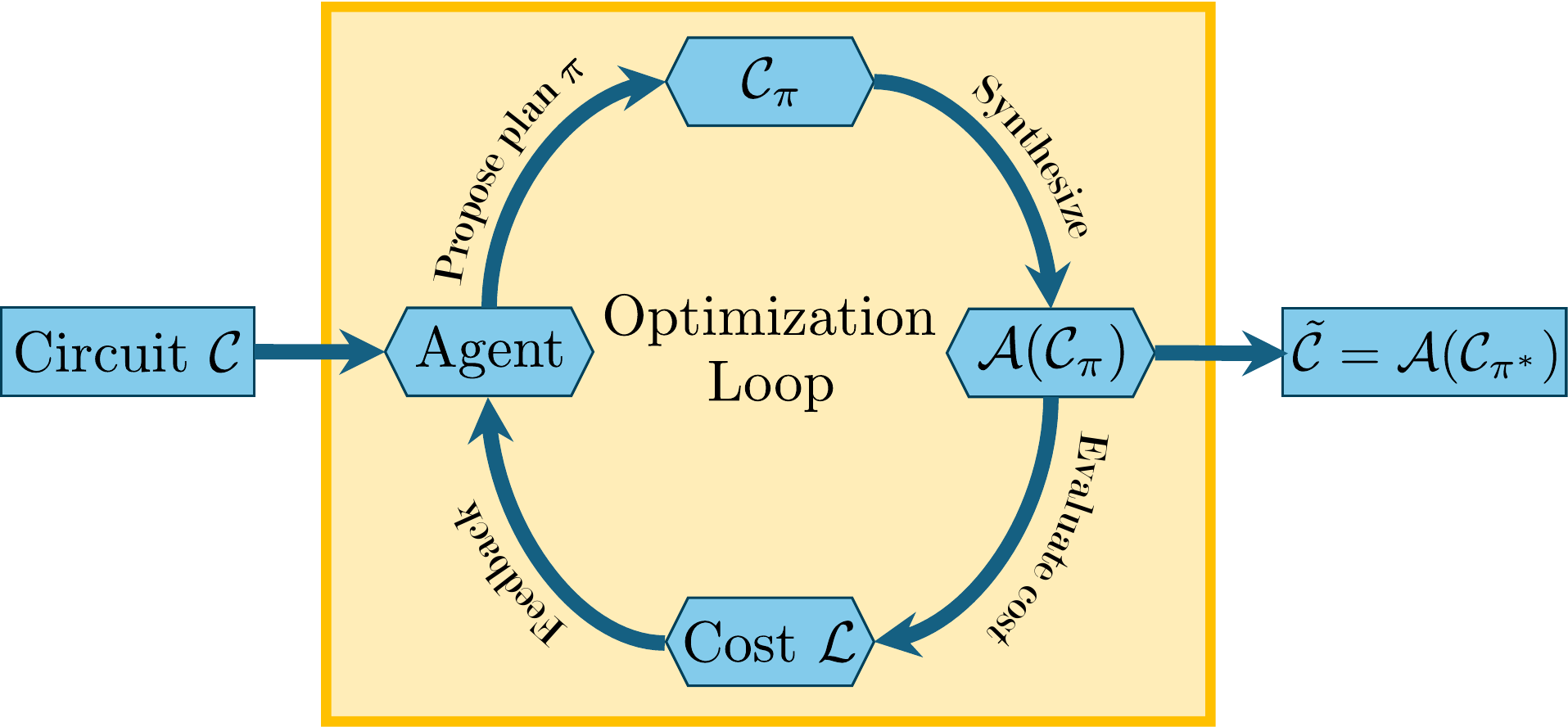}
    \caption{\textbf{Scheme of the proposed pipeline. }The RL agent proposes local edit operations on an input circuit \(C\), generating equivalent representations \(C_{\pi}\).
        Each modified circuit is compiled by a synthesis algorithm \(\mathcal{A}(C_{\pi})\) and evaluated according to a cost function \(\mathcal{L}\), typically reflecting the \(T\)-count.
        Feedback from this evaluation is used to update the agent’s policy, iteratively improving performance and yielding an optimized output circuit \(\tilde{C} = \mathcal{A}(C_{\pi^*})\). See Algorithm~\ref{alg:merge_synthesize} in the Methods Section for the pseudocode.   }
    \label{fig:pipeline-scheme}
\end{figure}

\subsubsection{Exploring the plan space}

The plan selection problem introduced above requires exploring a large combinatorial search space. 
Since the order of merges matters, even small circuits may admit a large number of possible plans. 
Exhaustive exploration is infeasible. Hence, we must instead rely on search strategies to efficiently 
identify promising sequences of actions. 
In this work, we investigate two strategies: a greedy baseline and a RL based approach. 

\paragraph{\textbf{Greedy search}.}

The greedy search strategy evaluates, at each step, all possible actions of the form 
\textsc{Two Qubit Merge}$(i,j)$. 
Each candidate action is applied temporarily to produce a modified circuit, 
which is then synthesized using $\cala$ and scored according to the cost function $\call(\cdot)$. 
The action yielding the largest immediate cost reduction is selected and applied. 
The procedure is then repeated on the updated circuit until no further improvements are found. While computationally straightforward, greedy search is inherently local: 
it may fail to identify plans where short-term cost increases are necessary to unlock 
larger reductions later in the sequence. 
Nevertheless, it provides a strong baseline and often produces substantial savings in practice. 

\paragraph{\textbf{Reinforcement learning search}.}

To overcome the limitations of greedy selection, we also implement an RL-based strategy. 
Here the process of plan construction is modeled as a sequential decision problem, 
where the \emph{state} corresponds to the current circuit representation, 
and the \emph{actions} are the possible merges $(i,j)$. 
At each timestep the agent selects an action according to a learned policy, 
with the goal of minimizing the final synthesis cost after a complete plan is executed.  The agent is trained with rewards derived from cost improvements, 
encouraging sequences of merges that lead to lower $T$-count after synthesis (see Fig.~\ref{fig:pipeline-scheme}). 
Unlike greedy search, the RL agent can learn to anticipate delayed rewards 
and avoid locally optimal but globally suboptimal choices. 
This allows it to navigate the plan space more effectively for larger circuits.

Our proposed agent is implemented using the Proximal Policy Optimization (PPO) algorithm \cite{PPO}, chosen for its stability and robustness in high-dimensional decision spaces through the use of clipped surrogate objectives. The policy is parameterized by a simple multilayer perceptron, which maps structured circuit representations to action probabilities. The observation space is defined as a three-channel tensor of dimension $n \times n \times 3$, where $n$ denotes the number of qubits. The first channel encodes, for each qubit pair $(i,j)$, the number of $T$ gates associated with wires $i$ and $j$. The second channel records the number of two-qubit gates that could be merged by selecting $(i,j)$ as the next action. For the final channel, we associate to each merge $(i,j)$ a binary matrix $M_{ij}$, defined such that the entries at positions $(i,i)$, $(j,j)$, and $(i,j)$ are set to $1$, while all other entries are $0$. The matrix representation of a complete plan is then obtained as the product of the matrices $M_{ij}$ corresponding to all merges in the plan. This formulation provides a compact encoding of the merge history and ensures invariance with respect to different merge orders that lead to equivalent overall plans.

The reward at each timestep is defined as the immediate improvement in $T$-count, i.e., the difference between the pre- and post-merge $T$-counts, thereby directly guiding the agent toward reductions in synthesis cost while enabling it to capture long-term dependencies across the plan space.

\paragraph{\textbf{Refining reinforcement learning plans}.}
Reinforcement learning is capable of exploring plans that exhibit long-range dependencies between actions, which are typically inaccessible to purely greedy strategies. However, because of the combinatorial size of the plan space and the inherent stochasticity of RL, there is no guarantee that an RL agent will identify an optimal plan. To address this limitation, a plan $\pi$ obtained from a non-greedy optimizer—such as an RL agent—can be further refined by evaluating, for each of its prefixes, the best greedy completion.

In detail, given a plan $\pi$, we perform greedy refinement by examining each prefix $\pi_i \subseteq \pi$ and selecting the action that yields the largest immediate improvement, iteratively extending the prefix until a complete plan is formed. This process produces, for each prefix, a greedily optimized completion, which in some cases leads to additional reductions in the $T$-count. At the end, the best plan is selected. Note that by construction, assuming that $\cala$ is deterministic, the plan $\pi^*$ obtained by greedy refinement from $\pi$ has a lower (or equal) $T$-count then both $\pi$, and the plan obtained by the greedy agent.

\subsubsection{Memoization}

In practice, applying $\cala$ repeatedly during search can be computationally expensive. 
Although different plans may generate different intermediate circuits, 
many of the local subcircuits encountered correspond to the same unitary blocks. To exploit this redundancy, we employ memoization: 
for each unitary $U$ passed to $\cala$, we cache its synthesized decomposition. 
Subsequent calls to $\cala(U)$ can then be served directly from the cache. 
This significantly reduces the computational cost of evaluating candidate actions, 
particularly in RL where many exploratory trajectories are generated.

\subsection{Experimental Setting}
\label{sec:exp}
To evaluate \methodname on different conditions, we select a variety of combinations of local synthesis algorithms and tasks. In this section we describe the pipeline implementation and the evaluation protocol.

\subsubsection{Local Synthesis Algorithms for Clifford+$T$}
\label{ssec:oralce}
In our context, a local synthesis algorithm $\cala$ is a procedure that, starting from a two-qubits unitary $U$, produces a decomposition $\cala(U)$ in a given set of gates $\mathcal{G}$. In practice, $\cala$ can be the composition of different algorithms, e.g., starting from a two-qubits unitary, we can first decompose it in CNOT+$U_3$ gates (where $U_3$ denotes a general unitary in $\mathcal{U}(2)$ parametrized by Euler's angles), and then decompose each $U_3$ gate in Clifford+$T$.

The "basic blocks" used in the experiments for Clifford+$T$ synthesis are:
\begin{itemize}
    \item \textit{TwoQubitBasisDecomposer}: implemented in Qiskit~\cite{qiskit}, starting from a two-qubits unitary $U$, it first decomposes $U$ with the KAK decomposition~\cite{drury_constructive_2008}, and then refines the decomposition with the Weyl chamber~\cite{Geometric_theory_nonLocal_op}. The resulting decomposition consists of up to $3$ CNOT gates, and up to $15$ single qubit rotations.
    \item \texttt{BQSKit}: The Berkeley Quantum Synthesis Toolkit~\cite{BQSKit} is a quantum compiler framework. When applied to synthesis with default settings, it uses an improvement of the approach proposed in~\cite{Quest} to produce a decomposition of a unitary $U$ in CNOT and $U_3$ gates, with the aim of minimizing the total number of resulting gates.
    \item \textit{Solovay Kitaev}: The SK algorithm~\cite{kitaev_quantum_1997, SK_alg} approximates arbitrary single-qubit unitaries using a finite universal gate set. It exploits the property that any unitary in $\mathcal{SU}(2)$ can be efficiently approximated to arbitrary precision $\varepsilon$.     In our setup, we use the Qiskit implementation of the SK algorithm with default arguments ($\{H, T, T^\dagger\}$ as the gate set and \texttt{recursion\_degree=2}).
    \item \textit{Gridsynth}: The \texttt{gridsynth} algorithm~\cite{gridsynth}, implemented in the \texttt{PyGridsynth} library, approximates single-qubit $R_z(\theta)$ rotations using the Clifford+$T$ gate set $\{H, S, T, X\}$ with a provably near-minimal number of $T$ gates. The method leverages number-theoretic techniques to achieve near-optimal synthesis with respect to both precision and $T$-count. In our experiments, arbitrary single-qubit unitaries are first decomposed into $U_3$ gates, i.e., three Euler rotations, each synthesized via \texttt{gridsynth}  with a fixed accuracy parameter $\varepsilon = 0.01$.
\end{itemize}

Each experiment performed on the task of Clifford+$T$ synthesis uses first a decomposition of a general unitary in CNOT and $U_3$ gates with either KAK decomposition or \texttt{BQSKit}, and then a synthesis of the resulting $U_3$ gates with either SK algorithm or \texttt{gridsynth}.

\subsubsection{Matchgate Synthesis}

To better highlight the generality of \methodname, we also evaluate the task of approximate matchgate synthesis~\cite{casas2025matchgate}. Here we recall that matchgates constitute a subset of unitaries in $\mathcal{U}(2^n)$ which can be obtained by single qubit rotations around the $z$ axis, and two qubit gates generate by $XX$ on neighboring qubits in a one-dimensional chain with open boundary conditions~\cite{matchgate_circuit}. Matchgates have received considerable interest as they constitute a classically simulable subset of circuits representing free fermionic evolutions. Recently, the work of~\cite{casas2025matchgate} proposed a technique synthesize matchgate circuits with matchgate-Clifford and $T$ gates.

The first step for synthesis is defining a denser gate set. As shown in~\cite{casas2025matchgate}, the set $\mathcal{G}= \{R_{xx}(\pi/2), \hat{S}, \hat{T}\}$, where $R_{xx}(\theta)=e^{-i  \theta XX/2}$ act on nearest neighbors in a one-dimensional chain with open boundary conditions, and where $\hat{S}=R_z(\pi/2)$ and $\hat{T}=R_z(\pi/4)$, with $R_{z}(\theta)=e^{-i \theta Z/2}$, is dense in matchgates, enabling for universal matchgate synthesis. Next, one starts from a matchgate unitary decomposed as $U=\prod_{l=1}^K U_l$, where we assume that each $U_l$ corresponds to either a $R_{xx}(\theta)$ or a $R_z(\theta)$~\cite{braccia_optimal_2025}. Then, we use the fact that the algebra generated by $X_i\otimes X_{i+1}$ and $Z_i$ is isomorphic to $\mathcal{SU}(2)$, meaning that  any gate $U_l$ can be transformed via some isomorphism $\phi$ into gate in the standard representation of $\mathcal{SU}(2)$, compiled there using standard gate synthesis tools (such as \texttt{gridsynth} or any of the methods previously discussed), and then transformed back into matchgates with $\phi^-1$. We refer the reader to~\cite{casas2025matchgate} for more details.  Ultimately, the key advantage of matchgate synthesis is that one leverages any standard compiler for local compilation in Clifford+$T$, meaning that all the tools developed for standard synthesis can be directly deployed for matchgates.

\subsubsection{RL Agent Implementation}

The training and evaluation procedure for the RL agent is implemented in Python with the library \texttt{StableBaseline3}~\cite{stable-baselines3}.
First, an environment of the task is provided starting from the target circuit $\calc$ (written in Qiskit) and the synthesis algorithm $\cala$. At each step, the agent selects an action as a \textsc{Two Qubit Merge(i,j)} between interacting qubits, and adds it to the current plan $\pi$. Then, the environment applies the plan $\pi$ to $\calc$, uses $\cala$ to synthesize the circuit and returns the synthesized circuit $\tilde{\calc}_\pi$ and the corresponding $T$-count. The reward is given by the difference between the previous $T$-count and the current one. The cumulative reward corresponds to the total $T$-count reduction with regard to the initial value.

The agent is modeled as a PPO algorithm~\cite{PPO} with the default settings provided by StableBaseline3, but with a \texttt{entropy\_coefficient} equal to $0.05$ to promote exploration. The agent is evaluated each $10^5$ steps, and the training is stopped after $10$ evaluations without performance increases.

To address the stochasticity of the plan selection, after the training the agent is evaluated multiple times. At the end, the plan with the lowest $T$-count is selected.

\subsection{Numerical implementation}
\label{sec:results}

\subsubsection{General Random Circuit}
\label{subsec:TASKA}

We consider the task of random circuit synthesis.
We fix the number of qubits $n\in\{4,6,8,10\}$, and we generate $5$ random circuits of $n$ qubits and depth $n$. Each circuit is generated with the Qiskit \texttt{random\_circuit} function. As each pair of qubits can interact, the number of actions is limited by $n(n-1)/2$.

To evaluate our approach for different synthesis algorithms, we selected the following:
\begin{itemize}
    \item \textit{KAK}$\times$\texttt{gridsynth};
    \item \texttt{BQSKit}$\times$\texttt{gridsynth};
    \item \textit{KAK}$\times$\textit{Solovay-Kitaev}.
\end{itemize}

\begin{figure}[t]
    \centering
\includegraphics[width=1\columnwidth]{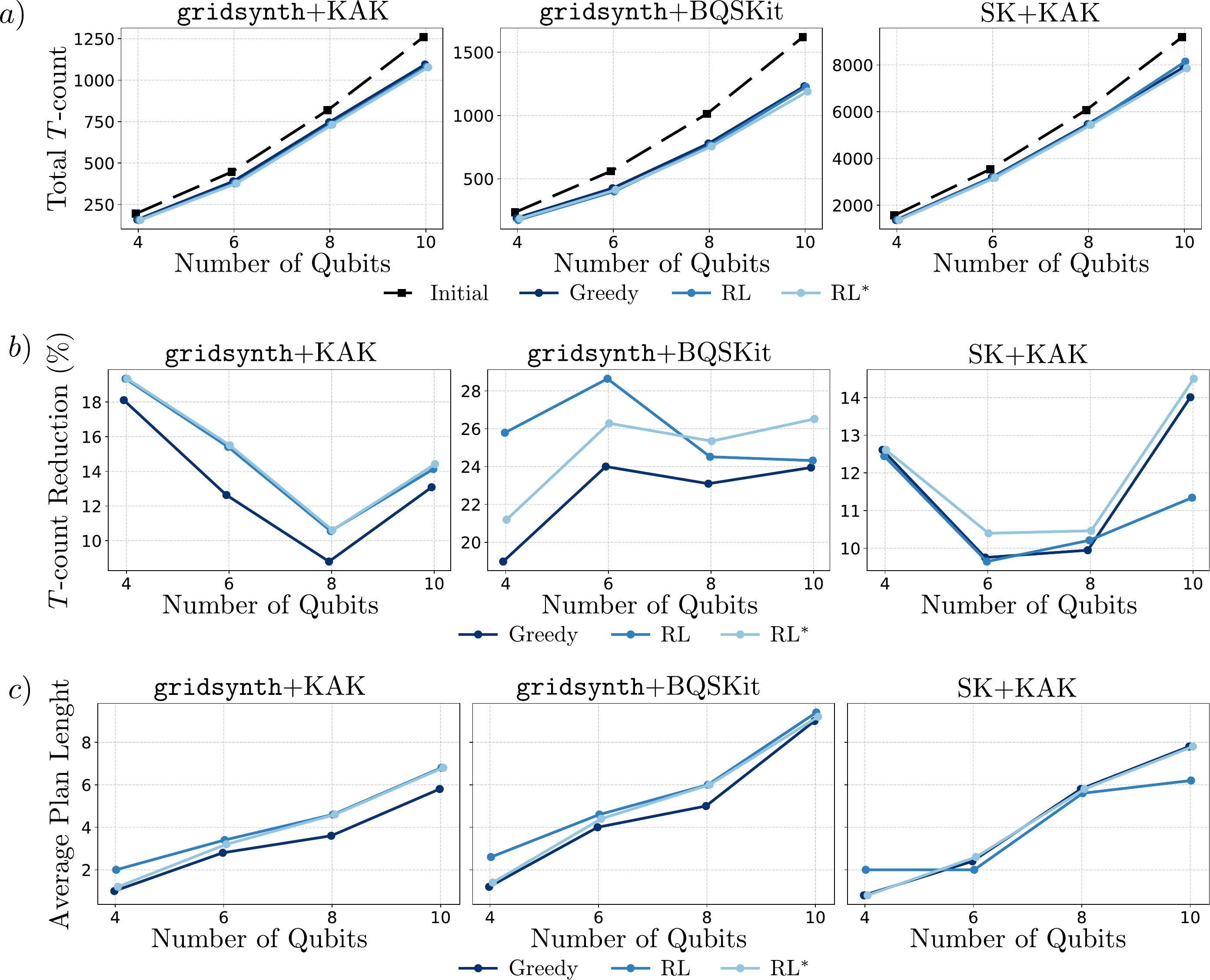}
    \caption{\textbf{Results for  random circuit synthesis} (Task \ref{subsec:TASKA}). Average of final $T$-count (a), $T$-count reduction in percentage (b), and plan length (c).}
    \label{fig:exp1}
\end{figure}

For each algorithm $\cala$, we evaluate the total number of $T$ gates obtained on average by the application of $\cala$ starting from the initial circuit representation, and the number of $T$ gates obtained after the application of the plan found by the Greedy, RL, and RL refined (denoted as RL$^*$) approaches. These results are shown in the Fig.~\ref{fig:exp1}(a). In addition, in Fig.~\ref{fig:exp1}(b) we show the reduction in percentage obtained by each agent with regard to the initial $T$-count. Figure~\ref{fig:exp1}(c) shows the average number of merges (i.e., the plan length).

\begin{figure}[t]
    \centering
\includegraphics[width=1\columnwidth]{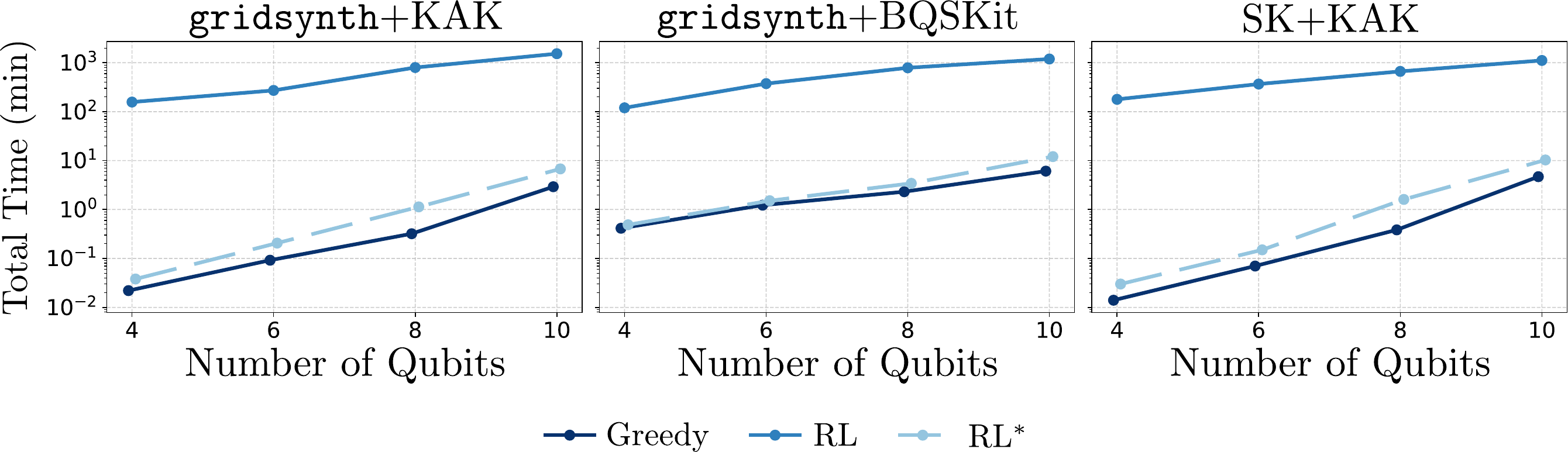}
    \caption{\textbf{Computational time for random circuit synthesis} (Task \ref{subsec:TASKA}). The dashed line, corresponding to the refinement time, is computed as the time required by the greedy refinement process after the training of the RL agent.}
    \label{fig:exp1time}
\end{figure}

As a first step, we observe that, in general, the final $T$-counts depend on the choice of $\cala$. With regard to initial $T$-counts (i.e., without pre-synthesis), \texttt{gridsynth}+KAK and \texttt{gridsynth}+\texttt{BQSKit} obtains similar results, with the former being slightly better than the latter. As expected, SK+KAK has a much worse scaling. In general, pre-synthesis allows a reduction of $T$-count between $10\%$ and $25\%$, depending on the choices of $\cala$ and on the search algorithm considered. 
In particular, algorithms based on KAK obtains a reduction between $10\%$ and $18\%$, while \texttt{gridsynth}+\texttt{BQSKit} obtained percentage reductions of more than $20\%$, with higher variability due to the high stochasticity of \texttt{BQSKit}.

With regards to the search strategy, we observe that Greedy and RL agents behave similarly, with RL obtaining almost always better reductions, and slightly longer plans, highlighting that greedy solutions are not optimal and more complex search strategies can be used to obtain better solutions. Note that even a difference of a few percentage points can be critical when the size of the system grows.

\begin{figure}[t]
    \centering
\includegraphics[width=1\columnwidth]{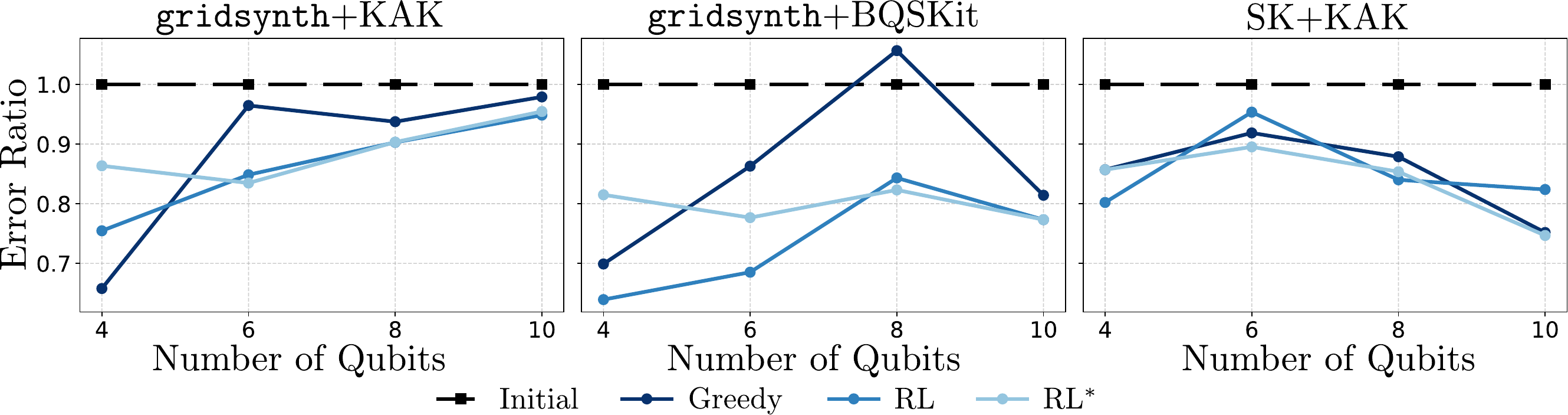}
    \caption{\textbf{Error for  random circuit synthesis} (Task \ref{subsec:TASKA}). Ration between the final approximation error obtained with no pre-synthesis, and the final approximation error after the application of the discovered plans.}
    \label{fig:exp1-err}
\end{figure}

\paragraph{Approximation Error}
As discussed in Section \ref{sec:method}, the application of pre-synthesis does not increase the total approximation error. In Fig.~\ref{fig:exp1-err} we show that, on contrary, it often allows to \textit{reduce} the total approximation error when compared to directly performing synthesis (fixed the approximation tolerance of $\cala$). This depends on the consideration that, at worst, approximation error grows additively with the number of starting gates. By merging sequences of gates we can reduce the number of total synthesis approximations, resulting in lower errors (compared to no pre-synthesis final circuits).

\begin{figure}[t]
    \centering
    \includegraphics[width=1\columnwidth]{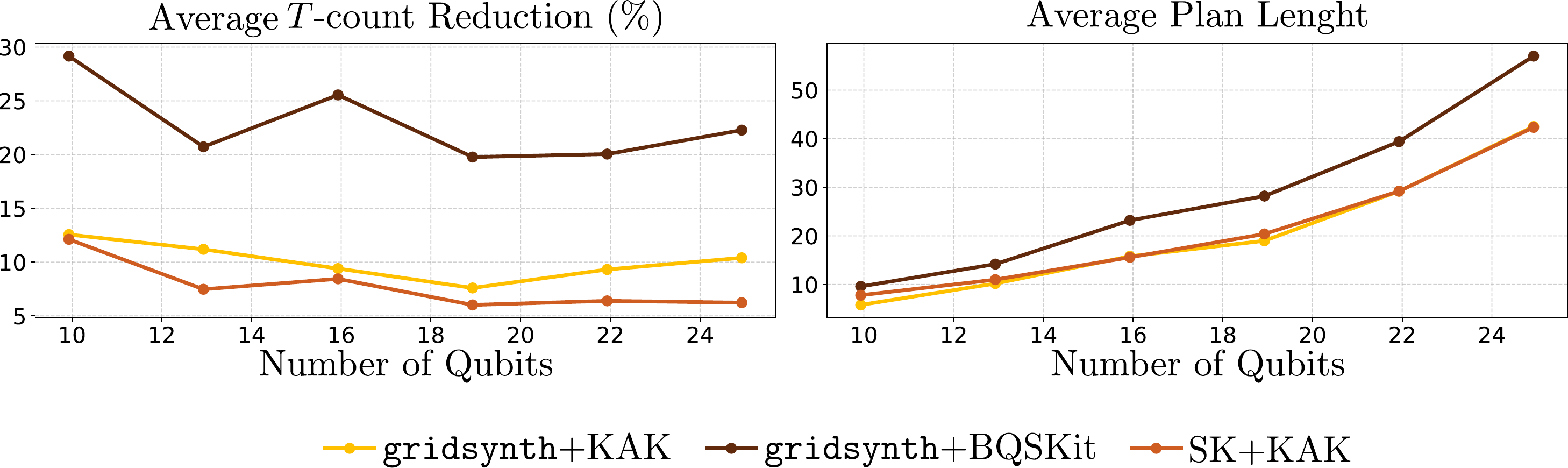}
    \caption{\textbf{Evaluation of $T$-count reduction and average plan length discovered by the greedy agent} (Task \ref{subsec:TASKA}(c)). We depict circuits on up-to $25$ qubits.}
    \label{fig:exp1-greedyscale}
\end{figure}

\paragraph{Scaling up to 25 qubits}
Finally, in Fig.~\ref{fig:exp1-greedyscale} we evaluate the performances on the greedy agent on random circuits of sizes up to 25 qubits. On smaller sizes we observed that RL and RL$^*$ are always able to obtain higher reductions than the greedy agent. On the other hand, training a RL agent is much more computationally expensive than using a greedy solution, suggesting a trade-off between computational time and performances. 

Results show that percentage $T$-count reductions are similar on different scales, obtaining an improvement of approximately $10\%$ on \texttt{gridsynth}+KAK, between $20\%$ and $25\%$ for \texttt{gridsynth}+\texttt{BQSKit}, and  between $5\%$ and $10\%$ for SK+KAK.   
We expect more complex search strategies to obtain higher $T$-count reductions, similarly to what we observed in on circuits of smaller size (Fig. \ref{fig:exp1}). However, as the size of plans space increases, better learning algorithms may be required (e.g. by combining RL with Monte Carlo Tree Search \cite{silver_mastering_2017, schrittwieser_mastering_2020}).

\paragraph{Computational Time:}\label{par:comp-time}
Finally, we briefly discuss the computational time required by this task. In general, the total time depends on the time to evaluate a plan (or a single action), times the number of evaluations. Note that given an action space of size $\ell$ (corresponding to the pairs of interacting qubits), the number of plans evaluated by the greedy algorithm is limited by $\ell^2$. On the other hand, a RL agent explore the plan space for a number of steps that usually depends on the convergence of the agent. In our case, each agent reached convergence between $10^6$ and $10^7$ steps. However, memoization on single qubit gates-level, two qubits gates-level, and plan-level allowed to reduce the total computational time by several orders of magnitude.

In practice, RL agents training required between 2 and 20 hours on average for a qubit counts between 4 and 10, while greedy agent required up to 10 minutes (Figure \ref{fig:exp1time}).
For qubit counts between 10 and 25, the greedy agent required up to 5 hours (Figure \ref{fig:exp1-greedyscale}, right).

\subsubsection{Linear Connectivity}
\label{subsec:TASKB}

\begin{figure}[t]
    \centering
\includegraphics[width=1\columnwidth]{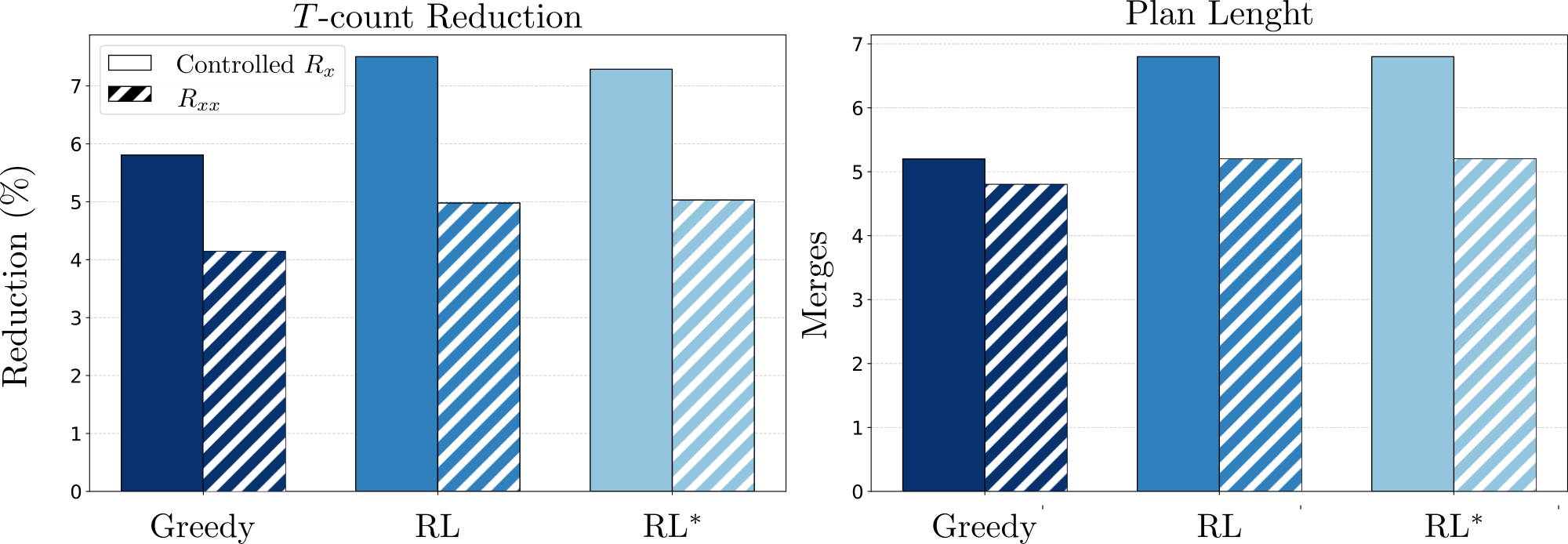}
    \caption{\textbf{Results on circuits with linear connectivity} (Task \ref{subsec:TASKB}). We show the reduction percentage for different approaches in terms of $T$-count and plan length.}
    \label{fig:exp2}
\end{figure}

We now consider circuits with linear connectivity. We build random circuits of 10 qubits as the concatenation of 3 blocks, where each block is composed by a $R_x$ and $R_z$ rotation on each qubit, followed by a linear entangling layer. We restrict our evaluation to \texttt{gridsynth}+KAK, as the results in Section \ref{subsec:TASKA} suggest it obtains generally lower $T$-count.
As the entangling layer, we consider two alternatives: a linear  ladder composed by controlled-$R_x$ gates, and a "brick-like" structure composed by alternating $R_{xx}$ rotations. We report the results in Fig.~\ref{fig:exp2}.

First, we observe that the two tasks (Controlled $R_x$ and $R_{xx}$ brick-like) has different $T$-count reductions and plan length.

As the number of interacting pairs of qubit is lower (up to 9 pairs, compared to the 45 with $10$ qubits and full connectivity in Section \ref{subsec:TASKA}), we would expect the average plan length to be smaller. However, the average length is close to 7 and 5 respectively, suggesting that most merges are useful to reduce the total $T$-count. However, due to the more regular structure, each merge has a smaller $T$-count reduction, compared to the setting with full connectvity.

This shows that on different settings, the pre-synthesis pipeline can provide different but consistent reductions.  

\subsubsection{Real-time Dynamics of the generalized Quantum Ising model}
\label{subsec:TASKC}

We next implement \methodname for the task of compiling real-time dynamics for the 1D Quantum Ising model. The Hamiltonian consists of a nearest-neighbour two-body interaction and a set of local fields,
\begin{equation}
    H = -\sum_i \big( J\, Z_i Z_{i+1} + h_x X_i + h_y Y_i + h_z Z_i \big),
\end{equation}
where the fields $h_x$, $h_y$, and $h_z$ are drawn identically and independently at random from $[-2,2]$ for each instance. v
To approximate the time-evolution operator $U(t) = e^{-iHt}$, we first compile a second-order Trotterization scheme with timestep $t/10$ via tensor networks using the method of Ref.~\cite{gibbs2025learning}. In a nutshell, the key idea of this technique is to take a short time evolution (which does not generate large amounts of entanglement), and compile it with tensor networks on an infinite matrix product states dataset~\cite{vidal2007classical} using tools from quantum machine learning to ensure good generalization~\cite{caro2021generalization,caro2022outofdistribution}. 

The ansatz for the tensor network compilation consists of layers of single qubit gates interleaved with two-qubit ones acting on nearest neighbors in a brick-like fashion. For this reason, we evaluate the compilation procedure in both \textit{globally}, and in layer-wise (LW) manner. That is, for each layer, we apply the pre-synthesis pipeline to reduce the number of $T$-gates independently, and then we combine them together. Finally, we note that due to symmetries in the circuits, many merges are equivalent up to translation of qubit indices, which makes it feasible to explore the relevant state space exhaustively for the considered system sizes. 

\begin{figure}[t]
    \centering
\includegraphics[width=0.99\columnwidth]{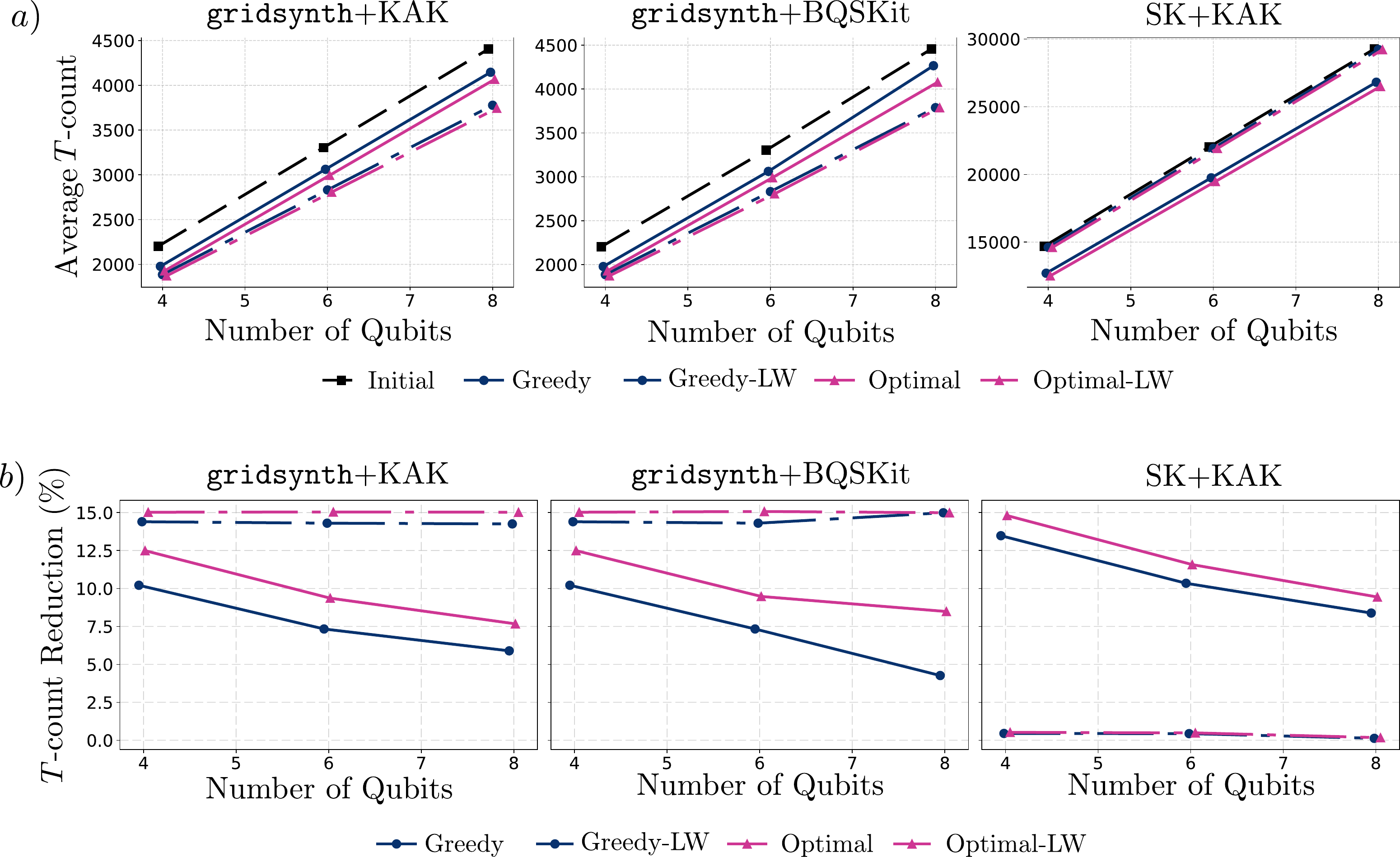}
    \caption{\textbf{Results on real-time dynamics compilation}  (Task \ref{subsec:TASKC}).  Average of final $T$-count (a), and $T$-count reduction in percentage (b). Optimal is obtained by brute-force search by exploiting the symmetries in the circuit structure. Dashed lines represent the counts for circuits that are optimized in a layer wise (LW) manner, where the circuit is first decomposed in layers, then each layer is optimized independently.}
    \label{fig:imps2}
\end{figure}

We report total $T$-counts and $T$-counts reductions in Figure \ref{fig:imps2}. We observe that the results for   \textit{KAK}$\times$\texttt{gridsynth} and \texttt{BQSKit}$\times$\texttt{gridsynth} are very similar. This follows from the fact that the structure of the merged two qubits gates offer no flexibility in the decomposition (as each merged gate arises from the combination of a $R_{ZZ}$ gate with two adjacent single qubit rotations). Hence, in this scenario, \textit{KAK} decomposition often produces the same result as \texttt{BQSKit}, leading to the same compiled circuit.

In general, we observe that the solutions obtained with \texttt{gridsynth} have a reduction close to $15\%$ for both greedy and optimal cases. Solutions found "globally" have a larger gate reduction, decreasing for higher number of qubits, and with a consistent gap in quality between the greedy plan and the optimal one. On the other hand, solutions for $SK$ are better when optimizing globally. This counter-intuitive results can be explained by noting that the layer-wise optimization does not take into account interaction between adjacent layers, which in turn, could be relevant depending on the synthesis algorithm considered.

\subsubsection{Matchgate Circuits}
\label{subsec:TASKD}

Finally, we consider the task of matchgate compilation. Starting from a random matchgate circuit $\calc$, composed by randomly selecting $R_z$ and $R_{xx}$ gates with local interactions, we aim to find a pre-synthesis strategy that reduces the total $T$-count when decomposing $\calc$ in the set $\{R_{xx}(\pi/2), \hat{S}, \hat{T}\}$.

For each number of qubits $n\in\{5, 10, 15, 20, 25\}$, we generate 10 matchgate circuits with $5n$ gates, and we use the greedy agent to explore the plan space.

\begin{figure}[t!]
    \centering
\includegraphics[width=1\columnwidth]{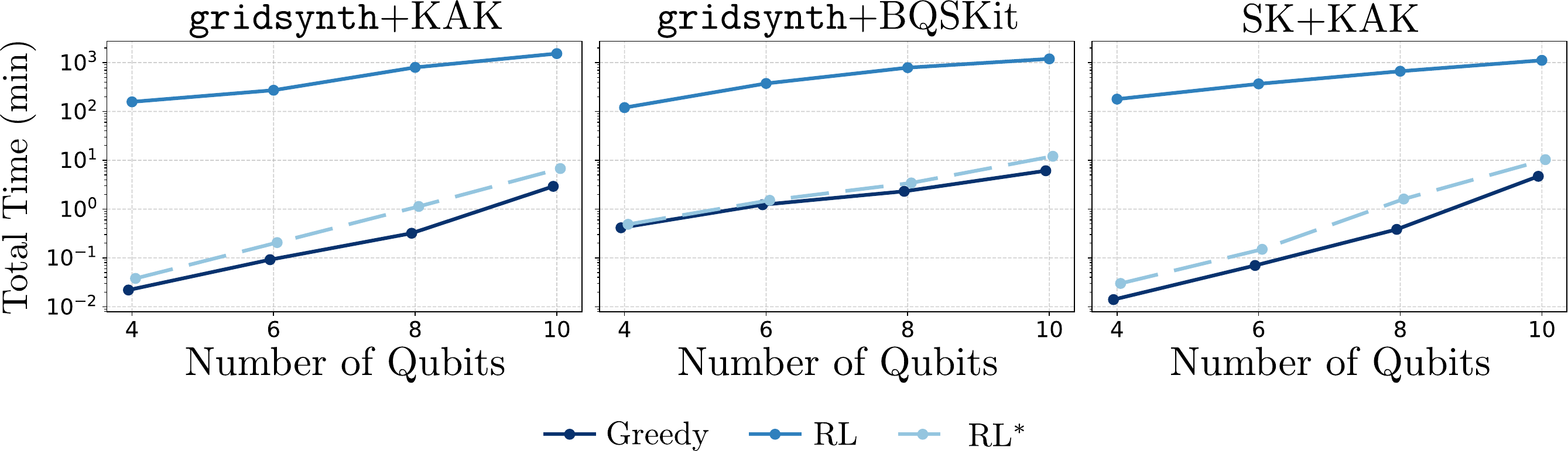}
    \caption{\textbf{Results on matchgate circuits} (Task \ref{subsec:TASKD}). We show the $T$-count and plan length reduction percentage for matchgate circuits on up-to 25 qubits, obtained by Greedy search.}
    \label{fig:exp3}
\end{figure}

We observe that, in contrast to other settings, the amount of merges considered is lower, suggesting again that both plan length and final reduction depends mostly on the initial circuit configuration, and synthesis approach. The final reduction in $T$-count is between $10\%$ and $15\%$, showing that our pre-synthesis pipeline can obtain a consistent $T$-count reduction in a variety of settings (see Fig.~\ref{fig:exp3}).

\section{Discussions}
\label{sec:conclusion}

We introduced a pre-synthesis framework, \methodname, that treats equivalent circuit representations as an optimizable design choice and showed that learning-guided local edits can substantially reduce the $T$-count produced by off-the-shelf local synthesis algorithms. Concretely, we defined unitary-preserving merge actions (single- and two-qubit merges), formalized plan selection as a sequential decision problem, and implemented both a greedy baseline and a reinforcement-learning PPO agent to explore the space of plans.

Our empirical evaluation across multiple synthesis algorithms demonstrates consistent reductions in post-synthesis $T$-count, with typical improvements in the range of 10–25\%. Plans obtained by the RL agent obtain further reduction compared to greedy solution, showing that better exploration strategies are beneficial for synthesis outcomes. These findings establish pre-synthesis as an effective, synthesis-agnostic preprocessing stage that can be applied on top of existing compilation pipelines to lower the cost of fault-tolerant implementations.

We close by highlighting important limitations and promising directions for future work. First, exploring merges acting on more than two qubits could unlock additional savings but requires more scalable synthesis or heuristic guidance due to the exponential growth of the search space.
Second, training RL agents for very large circuits remains computationally demanding; hybrid strategies that combine RL with planning algorithms (e.g., MCTS \cite{silver_mastering_2017, schrittwieser_mastering_2020}), curriculum learning \cite{narvekar2020curriculumlearningreinforcementlearning}, or transfer across related circuit families could improve scalability.
Finally, integrating topology awareness, cost-aware reward formulations, and a standardized benchmark suite will make it easier to compare pre-synthesis strategies and deploy them in end-to-end compilers.

In summary, representation-level optimization via learned local edits is a practical and broadly applicable technique for reducing non-native gate cost in fault-tolerant synthesis. We expect that combining the ideas presented here with more advanced search, larger merge kernels, and multi-function objectives (e.g., reducing both $T$-counts and circuit depth) will further increase the practical reach of pre-synthesis for near-term and future quantum devices.

\section{Methods}

\label{app:1}
We provide the pseudocode for the actions described in Section \ref{sec:method}. In particular, Algorithm \ref{alg:one_qubit_merge} describes the \textsc{Single Qubit Merge}, and Algorithm \ref{alg:two_qubit_merge_contiguous} the \textsc{Two Qubit Merge}.

The core idea is as follows:

\begin{itemize}
    \item \textsc{Single Qubit Merge:} For each qubit, maintain a buffer that accumulates consecutive single-qubit gates. When a gate acts on another qubit or a multi-qubit gate appears, the buffer is flushed by combining all accumulated gates into a single unitary, which is then appended to the output circuit.
    
    \item \textsc{Two Qubit Merge:} For a given qubit pair, maintain a buffer that accumulates contiguous 2-qubit gates acting only on that pair. If a gate touches one or both qubits but is not fully contained in the pair, the buffer is flushed into a single 2-qubit unitary. Single-qubit gates on either qubit are then merged into the neighboring 2-qubit unitaries.
\end{itemize}

Note that the pseudocode provided in Algorithm \ref{alg:two_qubit_merge_contiguous} can be easily extended for \textsc{K Qubit Merge}, as discussed in Section \ref{sec:method}.

\begin{algorithm}[htbp]
\caption{OneQubitMerge$(i)$}
\label{alg:one_qubit_merge}

\begin{algorithmic}[1]

\Require Circuit $\mathcal{C}$, target qubit index $i$
\Ensure Circuit $\mathcal{C}'$ with logically adjacent single-qubit gates on qubit $i$ merged

\State Initialize empty circuit $\mathcal{C}'$
\State Initialize empty buffer $B \gets []$ \Comment{Accumulator for qubit $i$}

\vspace{0.5em}
\Function{FlushBuffer}{}
    \If{$B$ is empty}
        \State \Return
    \EndIf
    \State Compute unitary $U$ of all gates in $B$ (in order)
    \If{$U$ is identity}
        \State Clear buffer $B \gets []$
        \State \Return
    \EndIf
    \State Decompose $U$ into a single $U3(\theta,\phi,\lambda)$ gate
    \State Append $U3(\theta,\phi,\lambda)$ to $\mathcal{C}'$ on qubit $i$
    \State Clear buffer $B \gets []$
\EndFunction
\vspace{0.5em}

\ForAll{instructions \textit{(instr, qubits)} in $\mathcal{C}$}
    \If{$instr$ acts only on qubit $i$}
        \State Append $instr$ to buffer $B$
    \Else
    \State $\textit{touched} \gets$ set of qubit indices in $\textit{qubits}$
        \If{$i \in \textit{touched}$ \textbf{or} $instr$ is multi-qubit}
            \State \Call{FlushBuffer}{}
        \EndIf
        \State Append \textit{(instr, qubits)} to $\mathcal{C}'$
    \EndIf
\EndFor

\State \Call{FlushBuffer}{} \Comment{Flush any remaining gates on qubit $i$}
\State \Return $\mathcal{C}'$

\end{algorithmic}
\end{algorithm}

Finally, in Algorithm \ref{alg:merge_synthesize} we describe the process to compute the $T$-count of a synthesized circuit $\calc$ given a local synthesis algorithm $\cala$ and a plan $\pi$, as it used in Equation \ref{eq:formula_pistar} and in the general scheme proposed in Figure  \ref{fig:pipeline-scheme}.

\section*{Acknowledgments}
We thank Paolo Braccia for providing us with unitaries for the matchgate compilation examples. L.C. and M.C. acknowledge support by the Laboratory Directed Research and Development (LDRD) program of Los Alamos National Laboratory (LANL) under project number 20260043DR and by  the U.S. Department of Energy, Office of Science, Office of Advanced Scientific Computing Research under Contract No. DE-AC05-00OR22725 through the Accelerated Research in Quatum Computing Program MACH-Q project. This work was also supported by the Quantum Science Center (QSC), a National Quantum Information Science Research Center of the U.S. Department of Energy (DOE).

\section{Data availability}

All code necessary to reproduce the results and evaluate the methods can be found at ~\cite{lizzio2025quantum}.

\section{Author contributions}

The project idea was proposed by DLB, GS and MC. The theoretical framework was developed by DLB and GS. Numerical simulations were performed by DLB and LC. The manuscript was written by DLB and MC. All authors reviewed the work.

\clearpage

\begin{algorithm}[htbp]
\caption{TwoQubitMerge$(i,j)$}
\label{alg:two_qubit_merge_contiguous}
\begin{algorithmic}[1]
\Require Circuit $\mathcal{C}$, qubit indices $(q_i, q_j)$
\Ensure Circuit $\mathcal{C}'$ with contiguous gates between $i$ and $j$ merged
\State Initialize empty circuit $\mathcal{C}'$
\State Initialize empty buffer $B \gets []$  \Comment{Temporary storage for contiguous 2-qubit gates on $(q_i, q_j)$}
\Statex
\Function{FlushBuffer}{}
    \If{$B$ is empty} \Return \EndIf
    \State Construct 2-qubit circuit $U$ from instructions in $B$ (map $q_i\mapsto0$, $q_j\mapsto1$)
    \State Append $U$ as a single 2-qubit unitary gate to $\mathcal{C}'$ on $(q_i, q_j)$
    \State Clear buffer $B$
\EndFunction
\Statex
\ForAll{instructions $(\textit{instr}, \textit{qubits})$ in $\mathcal{C}$}
    \State $\textit{touched} \gets$ set of qubit indices in $\textit{qubits}$
    \If{$\textit{touched} \subseteq \{q_i, q_j\}$}
        \State Append $(\textit{instr}, \textit{qubits})$ to buffer $B$ \Comment{Accumulate contiguous 2-qubit gates}
    \ElsIf{$\textit{touched} \cap \{q_i, q_j\} \neq \emptyset$}
        \State \Call{FlushBuffer}{}
        \State Append $(\textit{instr}, \textit{qubits})$ to $\mathcal{C}'$
    \Else
        \State Append $(\textit{instr}, \textit{qubits})$ to $\mathcal{C}'$
    \EndIf
\EndFor
\State \Call{FlushBuffer}{}  \Comment{Flush remaining gates in buffer}

\State $\mathcal{C}' \gets \Call{OneQubitMergeIntoTwoQ}{\mathcal{C}', i, j, i}$  \Comment{Merge 1-qubit gates on $q_i$ into 2-qubit gates}
\State $\mathcal{C}' \gets \Call{OneQubitMergeIntoTwoQ}{\mathcal{C}', i, j, j}$  \Comment{Merge 1-qubit gates on $q_j$ into 2-qubit gates}

\State \Return $\mathcal{C}'$
\end{algorithmic}
\end{algorithm}

\begin{algorithm}[htbp]
\caption{Merge and Synthesize}
\label{alg:merge_synthesize}
\begin{algorithmic}[1]
\Require Circuit $\mathcal{C}$, local synthesis algorithm $\mathcal{A}$, merge plan $\pi$
\Ensure Clifford+$T$ circuit $\widetilde{\mathcal{C}}$ and total $T$-count $T_{\mathrm{count}}$
\Statex

\State $\mathcal{C} \gets \Call{RemoveIdentities}{\mathcal{C}}$ \Comment{Eliminate redundant identity operations}

\ForAll{$(i,j) \in \pi$}
    \State $\mathcal{C} \gets \Call{TwoQubitMerge}{\mathcal{C}, (i,j)}$
\EndFor

\ForAll{two-qubit unitaries $U \in \mathcal{C}$}
    \State Replace $U$ with $\mathcal{A}(U)$ \Comment{Apply local two-qubit synthesis}
\EndFor

\If{single-qubit merging is enabled}
    \For{$i = 1$ \textbf{to} $n$}
        \State $\mathcal{C} \gets \Call{OneQubitMerge}{\mathcal{C}, i}$
    \EndFor

\ForAll{one-qubit unitaries $U \in \mathcal{C}$}
    \State Replace $U$ with $\mathcal{A}(U)$ \Comment{Apply local single-qubit synthesis}
\EndFor
\EndIf

\State $T_{\mathrm{count}} \gets \Call{CountT}{\mathcal{C}}$
\State \Return $\widetilde{\mathcal{C}}, T_{\mathrm{count}}$
\end{algorithmic}
\end{algorithm}

\clearpage
\newpage

\bibliographystyle{unsrtnat}
\bibliography{bib}

@inproceedings{davis_towards_2020,
	address = {Denver, CO, USA},
	title = {Towards {Optimal} {Topology} {Aware} {Quantum} {Circuit} {Synthesis}},
	copyright = {https://ieeexplore.ieee.org/Xplorehelp/downloads/license-information/IEEE.html},
	isbn = {9781728189697},
	url = {https://ieeexplore.ieee.org/document/9259942/},
	doi = {10.1109/QCE49297.2020.00036},
	urldate = {2025-10-23},
	booktitle = {2020 {IEEE} {International} {Conference} on {Quantum} {Computing} and {Engineering} ({QCE})},
	publisher = {IEEE},
	author = {Davis, Marc G. and Smith, Ethan and Tudor, Ana and Sen, Koushik and Siddiqi, Irfan and Iancu, Costin},
	month = oct,
	year = {2020},
	pages = {223--234},
}

@article{he_unitary_2023,
	title = {Unitary {Diagonalization} of the {Generalized} {Complementary} {Covariance} {Quaternion} {Matrices} with {Application} in {Signal} {Processing}},
	volume = {11},
	issn = {2227-7390},
	url = {https://www.mdpi.com/2227-7390/11/23/4840},
	doi = {10.3390/math11234840},
	language = {en},
	number = {23},
	urldate = {2025-10-23},
	journal = {Mathematics},
	author = {He, Zhuo-Heng and Zhang, Xiao-Na and Chen, Xiaojing},
	month = dec,
	year = {2023},
	pages = {4840},
}

@article{amy_meet_middle_2013,
	title = {A {Meet}-in-the-{Middle} {Algorithm} for {Fast} {Synthesis} of {Depth}-{Optimal} {Quantum} {Circuits}},
	volume = {32},
	copyright = {https://ieeexplore.ieee.org/Xplorehelp/downloads/license-information/IEEE.html},
	issn = {0278-0070, 1937-4151},
	url = {http://ieeexplore.ieee.org/document/6516700/},
	doi = {10.1109/TCAD.2013.2244643},
	number = {6},
	urldate = {2025-10-23},
	journal = {IEEE Transactions on Computer-Aided Design of Integrated Circuits and Systems},
	author = {Amy, M. and Maslov, D. and Mosca, M. and Roetteler, M.},
	month = jun,
	year = {2013},
	pages = {818--830},
}

@article{kliuchnikov_fast_2013,
	title = {Fast and efficient exact synthesis of single-qubit unitaries generated by {Clifford} and {T} gates},
	volume = {13},
	issn = {15337146, 15337146},
	url = {http://www.rintonpress.com/journals/doi/QIC13.7-8-4.html},
	doi = {10.26421/QIC13.7-8-4},
	number = {7\&8},
	urldate = {2025-10-23},
	journal = {Quantum Information and Computation},
	author = {Kliuchnikov, Vadym and Maslov, Dmitri and Mosca, Michele},
	month = may,
	year = {2013},
	pages = {607--630},
}

@article{gouzien2025provably,
  title={Provably optimal exact gate synthesis from a discrete gate set},
  author={Gouzien, {\'E}lie and Sangouard, Nicolas},
  journal={arXiv preprint arXiv:2503.15452},
  year={2025},
  doi={10.48550/arXiv.2503.15452},
  url={https://arxiv.org/abs/2503.15452}
}

@article{tang2019qubit,
  title={qubit-adapt-vqe: An adaptive algorithm for constructing hardware-efficient ans{\"a}tze on a quantum processor},
  author={Tang, Ho Lun and Shkolnikov, VO and Barron, George S and Grimsley, Harper R and Mayhall, Nicholas J and Barnes, Edwin and Economou, Sophia E},
  journal={PRX Quantum},
  volume={2},
  number={2},
  pages={020310},
  year={2021},
  publisher={APS},
  doi={10.1103/PRXQuantum.2.020310},
url={https://journals.aps.org/prxquantum/abstract/10.1103/PRXQuantum.2.020310}
}

@article{grimsley2019adaptive,
  title={An adaptive variational algorithm for exact molecular simulations on a quantum computer},
  author={Grimsley, Harper R and Economou, Sophia E and Barnes, Edwin and Mayhall, Nicholas J},
  journal={Nature {C}ommunications},
  volume={10},
  number={1},
  pages={1--9},
  year={2019},
  publisher={Nature Publishing Group},
  url={https://www.nature.com/articles/s41467-019-10988-2},
  doi={10.1038/s41467-019-10988-2}
}

@article{cincio2018learning,
	doi = {10.1088/1367-2630/aae94a},
	url = {https://doi.org/10.1088%2F1367-2630%2Faae94a},
	year = 2018,
	month = {nov},
	publisher = {{IOP} Publishing},
	volume = {20},
	number = {11},
	pages = {113022},
	author = {Lukasz Cincio and Yi{\u{g}}it Suba{\c{s}}{\i} and Andrew T Sornborger and Patrick J Coles},
	title = {Learning the quantum algorithm for state overlap},
	journal = {New Journal of Physics},
}

@article{moro2021quantum,
	author = {Moro, Lorenzo and Paris, Matteo G. A. and Restelli, Marcello and Prati, Enrico},
	date = {2021/08/06},
	doi = {10.1038/s42005-021-00684-3},
	id = {Moro2021},
	isbn = {2399-3650},
	journal = {Communications Physics},
	number = {1},
	pages = {178},
	title = {Quantum compiling by deep reinforcement learning},
	url = {https://doi.org/10.1038/s42005-021-00684-3},
	volume = {4},
	year = {2021}}

@article{du2020quantum,
  title={Quantum circuit architecture search: error mitigation and trainability enhancement for variational quantum solvers},
  author={Du, Yuxuan and Huang, Tao and You, Shan and Hsieh, Min-Hsiu and Tao, Dacheng},
  journal={npj Quantum Information},
  volume={8},
  number={1},
  pages={62},
  year={2022},
  publisher={Nature Publishing Group UK London},
  url={https://www.nature.com/articles/s41534-022-00570-y},
  doi={10.1038/s41534-022-00570-y}
}

@article{sim2021adaptive,
	doi = {10.1088/2058-9565/abe107},
	url = {https://doi.org/10.1088/2058-9565/abe107},
	year = 2021,
	month = {mar},
	publisher = {{IOP} Publishing},
	volume = {6},
	number = {2},
	pages = {025019},
	author = {Sukin Sim and Jonathan Romero and J{\'{e}}r{\^{o}}me F Gonthier and Alexander A Kunitsa},
	title = {Adaptive pruning-based optimization of parameterized quantum circuits},
	journal = {Quantum Science and Technology},
}

@article{zhang2021mutual,
  title={Mutual information-assisted adaptive variational quantum eigensolver},
  author={Zhang, Zi-Jian and Kyaw, Thi Ha and Kottmann, Jakob and Degroote, Matthias and Aspuru-Guzik, Alan},
  journal={Quantum Science and Technology},
  year={2021},
  publisher={IOP Publishing},
  url={https://iopscience.iop.org/article/10.1088/2058-9565/abdca4},
  doi={10.1088/2058-9565/abdca4}
}

@article{tkachenko2020correlation,
  title={Correlation-informed permutation of qubits for reducing ansatz depth in vqe},
  author={Tkachenko, Nikolay V and Sud, James and Zhang, Yu and Tretiak, Sergei and Anisimov, Petr M and Arrasmith, Andrew T and Coles, Patrick J. and Cincio, Lukasz and Dub, Pavel A},
  journal={PRX Quantum},
  volume={2},
  number={2},
  pages={020337},
  year={2021},
  publisher={APS},
  doi = {10.1103/PRXQuantum.2.020337},
  url={https://journals.aps.org/prxquantum/abstract/10.1103/PRXQuantum.2.020337}
}

@article{claudino2020benchmarking,
  title={Benchmarking adaptive variational quantum eigensolvers},
  author={Claudino, Daniel and Wright, Jerimiah and McCaskey, Alexander J and Humble, Travis S},
  journal={Frontiers in Chemistry},
  volume={8},
  pages={1152},
  year={2020},
  publisher={Frontiers},
  url={https://www.frontiersin.org/articles/10.3389/fchem.2020.606863/full},
  doi={10.3389/fchem.2020.606863}
}

@article{rattew2019domain,
  title={A domain-agnostic, noise-resistant, hardware-efficient evolutionary variational quantum eigensolver},
  author={Rattew, Arthur G and Hu, Shaohan and Pistoia, Marco and Chen, Richard and Wood, Steve},
  journal={arXiv preprint arXiv:1910.09694},
  year={2019},
  url={https://arxiv.org/abs/1910.09694}, 
doi={10.48550/arXiv.1910.09694}
}

@article{chivilikhin2020mog,
  title={MoG-VQE: Multiobjective genetic variational quantum eigensolver},
  author={Chivilikhin, D and Samarin, A and Ulyantsev, V and Iorsh, I and Oganov, AR and Kyriienko, O},
  journal={arXiv preprint arXiv:2007.04424},
  year={2020},
  url={https://arxiv.org/abs/2007.04424}, 
doi={10.48550/arXiv.2007.04424}
}

@article{zhang2020differentiable,
  title={Differentiable quantum architecture search},
  author={Zhang, Shi-Xin and Hsieh, Chang-Yu and Zhang, Shengyu and Yao, Hong},
  journal={Quantum Science and Technology},
  volume={7},
  number={4},
  pages={045023},
  year={2022},
  publisher={IOP Publishing},
url={https://iopscience.iop.org/article/10.1088/2058-9565/ac87cd},
doi={10.1088/2058-9565/ac87cd}
}

@article{wada2022sequential,
  title={Sequential optimal selections of single-qubit gates in parameterized quantum circuits},
  author={Wada, Kaito and Raymond, Rudy and Sato, Yuki and Watanabe, Hiroshi C},
  journal={Quantum Science and Technology},
  volume={9},
  number={3},
  pages={035030},
  year={2024},
  publisher={IOP Publishing}, 
url={https://iopscience.iop.org/article/10.1088/2058-9565/ad4583}, 
doi={10.1088/2058-9565/ad4583}
}

@article{li2024quarl,
  title={Quarl: A learning-based quantum circuit optimizer},
  author={Li, Zikun and Peng, Jinjun and Mei, Yixuan and Lin, Sina and Wu, Yi and Padon, Oded and Jia, Zhihao},
  journal={Proceedings of the ACM on Programming Languages},
  volume={8},
  number={OOPSLA1},
  pages={555--582},
  year={2024},
  publisher={ACM New York, NY, USA},
url={https://dl.acm.org/doi/10.1145/3649831},
doi={10.1145/3649831}
}

@article{fosel2021quantum,
  title={Quantum circuit optimization with deep reinforcement learning},
  author={F{\"o}sel, Thomas and Niu, Murphy Yuezhen and Marquardt, Florian and Li, Li},
  journal={arXiv preprint arXiv:2103.07585},
  year={2021},
  url={https://arxiv.org/abs/2103.07585},
  doi={10.48550/arXiv.2103.07585
}
}

@article{nakaji2025quantum,
  title={Quantum circuits as a game: A reinforcement learning agent for quantum compilation and its application to reconfigurable neutral atom arrays},
  author={Nakaji, Kouhei and Wurtz, Jonathan and Huang, Haozhe and Calder{\'o}n, Luis Mantilla and Panicker, Karthik and Kyoseva, Elica and Aspuru-Guzik, Al{\'a}n},
  journal={arXiv preprint arXiv:2506.05536},
  year={2025},
url={https://arxiv.org/abs/2506.05536},
doi={10.48550/arXiv.2506.05536}
}

@article{rosenhahn2023monte,
  title={Monte Carlo graph search for quantum circuit optimization},
  author={Rosenhahn, Bodo and Osborne, Tobias J},
  journal={Physical Review A},
  volume={108},
  number={6},
  pages={062615},
  year={2023},
  publisher={APS},
url={https://journals.aps.org/pra/abstract/10.1103/PhysRevA.108.062615},
doi={10.1103/PhysRevA.108.062615}
}

@book{nielsen2000quantum,
 author = {Michael A. Nielsen and Isaac L. Chuang},
 year = {2000},
 title = {Quantum Computation and Quantum Information},
 publisher = {Cambridge University Press},
 address = {Cambridge}
}

@article{kitaev_quantum_1997,
	title = {Quantum computations: algorithms and error correction},
	volume = {52},
	issn = {0036-0279, 1468-4829},
	shorttitle = {Quantum computations},
	url = {https://www.mathnet.ru/eng/rm892},
	doi = {10.1070/RM1997v052n06ABEH002155},
	number = {6},
	urldate = {2025-10-23},
	journal = {Russian Mathematical Surveys},
	author = {Kitaev, A Yu},
	month = dec,
	year = {1997},
	pages = {1191--1249},
}

@article{kukliansky2024leveraging,
  title={Leveraging Quantum Machine Learning Generalization to Significantly Speed-up Quantum Compilation},
  author={Kukliansky, Alon and Cincio, Lukasz and Younis, Ed and Iancu, Costin},
  journal={arXiv preprint arXiv:2405.12866},
  year={2024},
  doi={10.48550/arXiv.2405.12866}, 
  url={https://arxiv.org/abs/2405.12866}
}

@article{zhang2024scalable,
  title={Scalable quantum dynamics compilation via quantum machine learning},
  author={Zhang, Yuxuan and Wiersema, Roeland and Carrasquilla, Juan and Cincio, Lukasz and Kim, Yong Baek},
  journal={arXiv preprint arXiv:2409.16346},
  year={2024},
  doi={10.48550/arXiv.2409.16346}, 
  url={https://arxiv.org/abs/2409.16346}
}

@article{gridsynth,
author = {Ross, Neil J. and Selinger, Peter},
title = {Optimal ancilla-free Clifford+T approximation of z-rotations},
year = {2016},
issue_date = {September 2016},
publisher = {Rinton Press, Incorporated},
address = {Paramus, NJ},
volume = {16},
number = {11–12},
issn = {1533-7146},
journal = {Quantum Info. Comput.},
month = sep,
pages = {901–953},
numpages = {53},
url={https://dl.acm.org/doi/abs/10.5555/3179330.3179331}
}

@article{SK_alg,
author = {Dawson, Christopher M. and Nielsen, Michael A.},
title = {The Solovay-Kitaev algorithm},
year = {2006},
issue_date = {January 2006},
publisher = {Rinton Press, Incorporated},
address = {Paramus, NJ},
volume = {6},
number = {1},
issn = {1533-7146},
journal = {Quantum Info. Comput.},
month = jan,
pages = {81–95},
numpages = {15}
}

@article{Geometric_theory_nonLocal_op,
  title = {Geometric theory of nonlocal two-qubit operations},
  author = {Zhang, Jun and Vala, Jiri and Sastry, Shankar and Whaley, K. Birgitta},
  journal = {Phys. Rev. A},
  volume = {67},
  issue = {4},
  pages = {042313},
  numpages = {18},
  year = {2003},
  month = {Apr},
  publisher = {American Physical Society},
  doi = {10.1103/PhysRevA.67.042313},
  url = {https://link.aps.org/doi/10.1103/PhysRevA.67.042313}
}

@article{alphatensor_quantum,
      author={Ruiz, Francisco J. R. and Laakkonen, Tuomas and Bausch, Johannes and Balog, Matej and Barekatain, Mohammadamin and Heras, Francisco J. H. and Novikov, Alexander and Fitzpatrick, Nathan and Romera-Paredes, Bernardino and van de Wetering, John and Fawzi, Alhussein and Meichanetzidis, Konstantinos and Kohli, Pushmeet},
      title={Quantum Circuit Optimization with {A}lpha{T}ensor},
      journal = {Nature Machine Intelligence},
      year={2025},
      doi={10.1038/s42256-025-01001-1},
      url={https://www.nature.com/articles/s42256-025-01001-1},
}

@article{Bravyi2021cliffordcircuit,
  doi = {10.22331/q-2021-11-16-580},
  url = {https://doi.org/10.22331/q-2021-11-16-580},
  title = {Clifford {C}ircuit {O}ptimization with {T}emplates and {S}ymbolic {P}auli {G}ates},
  author = {Bravyi, Sergey and Shaydulin, Ruslan and Hu, Shaohan and Maslov, Dmitri},
  journal = {{Quantum}},
  issn = {2521-327X},
  publisher = {{Verein zur F{\"{o}}rderung des Open Access Publizierens in den Quantenwissenschaften}},
  volume = {5},
  pages = {580},
  month = nov,
  year = {2021}
}

@article{datastructure,
author = {Prasad, Aditya K. and Shende, Vivek V. and Markov, Igor L. and Hayes, John P. and Patel, Ketan N.},
title = {Data structures and algorithms for simplifying reversible circuits},
year = {2006},
issue_date = {October 2006},
publisher = {Association for Computing Machinery},
address = {New York, NY, USA},
volume = {2},
number = {4},
issn = {1550-4832},
url = {https://doi.org/10.1145/1216396.1216399},
doi = {10.1145/1216396.1216399},
journal = {J. Emerg. Technol. Comput. Syst.},
month = oct,
pages = {277–293},
numpages = {17}
}

@INPROCEEDINGS{QGo,
  author={Wu, Xin-Chuan and Davis, Marc Grau and Chong, Frederic T. and Iancu, Costin},
  booktitle={2021 International Conference on Rebooting Computing (ICRC)}, 
  title={Reoptimization of Quantum Circuits via Hierarchical Synthesis}, 
  year={2021},
  volume={},
  number={},
  pages={35-46},
  doi={10.1109/ICRC53822.2021.00016},
url={https://ieeexplore.ieee.org/document/9743148/}}

@INPROCEEDINGS{top-as,
  author={Weiden, Mathias and Kalloor, Justin and Kubiatowicz, John and Younis, Ed and Iancu, Costin},
  booktitle={2022 IEEE/ACM Third International Workshop on Quantum Computing Software (QCS)}, 
  title={Wide Quantum Circuit Optimization with Topology Aware Synthesis}, 
  year={2022},
  volume={},
  number={},
  pages={1-11},
  doi={10.1109/QCS56647.2022.00006},
url={https://ieeexplore.ieee.org/document/10025533}}

@inproceedings{Quest,
author = {Patel, Tirthak and Younis, Ed and Iancu, Costin and de Jong, Wibe and Tiwari, Devesh},
title = {QUEST: systematically approximating Quantum circuits for higher output fidelity},
year = {2022},
isbn = {9781450392051},
publisher = {Association for Computing Machinery},
address = {New York, NY, USA},
url = {https://doi.org/10.1145/3503222.3507739},
doi = {10.1145/3503222.3507739},
booktitle = {Proceedings of the 27th ACM International Conference on Architectural Support for Programming Languages and Operating Systems},
pages = {514–528},
numpages = {15},
location = {Lausanne, Switzerland},
series = {ASPLOS '22}
}

@INPROCEEDINGS{Quest2,
  author={Clark, Joseph and Thapliyal, Himanshu},
  booktitle={2024 25th International Symposium on Quality Electronic Design (ISQED)}, 
  title={Peephole Optimization for Quantum Approximate Synthesis}, 
  year={2024},
  volume={},
  number={},
  pages={1-8},
  doi={10.1109/ISQED60706.2024.10528701},
url={https://ieeexplore.ieee.org/document/10528701}}

@article{Quarl,
author = {Li, Zikun and Peng, Jinjun and Mei, Yixuan and Lin, Sina and Wu, Yi and Padon, Oded and Jia, Zhihao},
title = {Quarl: A Learning-Based Quantum Circuit Optimizer},
year = {2024},
issue_date = {April 2024},
publisher = {Association for Computing Machinery},
address = {New York, NY, USA},
volume = {8},
number = {OOPSLA1},
url = {https://doi.org/10.1145/3649831},
doi = {10.1145/3649831},
journal = {Proc. ACM Program. Lang.},
month = apr,
articleno = {114},
numpages = {28}
}

@article{RLxZXC,
  doi = {10.22331/q-2025-05-28-1758},
  url = {https://doi.org/10.22331/q-2025-05-28-1758},
  title = {Reinforcement {L}earning {B}ased {Q}uantum {C}ircuit {O}ptimization via {ZX}-{C}alculus},
  author = {Riu, Jordi and Nogu{\'{e}}, Jan and Vilaplana, Gerard and Garcia-Saez, Artur and Estarellas, Marta P.},
  journal = {{Quantum}},
  issn = {2521-327X},
  publisher = {{Verein zur F{\"{o}}rderung des Open Access Publizierens in den Quantenwissenschaften}},
  volume = {9},
  pages = {1758},
  month = may,
  year = {2025}
}

@misc{QCGame,
      title={Quantum circuits as a game: A reinforcement learning agent for quantum compilation and its application to reconfigurable neutral atom arrays}, 
      author={Kouhei Nakaji and Jonathan Wurtz and Haozhe Huang and Luis Mantilla Calderón and Karthik Panicker and Elica Kyoseva and Alán Aspuru-Guzik},
      year={2025},
      eprint={2506.05536},
      archivePrefix={arXiv},
      primaryClass={quant-ph},
      url={https://arxiv.org/abs/2506.05536}, 
}

@misc{fösel2021quantumcircuitoptimizationdeep,
      title={Quantum circuit optimization with deep reinforcement learning}, 
      author={Thomas Fösel and Murphy Yuezhen Niu and Florian Marquardt and Li Li},
      year={2021},
      eprint={2103.07585},
      archivePrefix={arXiv},
      primaryClass={quant-ph},
      url={https://arxiv.org/abs/2103.07585}, 
}

@misc{wang2024quantumcompilingreinforcementlearning,
      title={Quantum Compiling with Reinforcement Learning on a Superconducting Processor}, 
      author={Z. T. Wang and Qiuhao Chen and Yuxuan Du and Z. H. Yang and Xiaoxia Cai and Kaixuan Huang and Jingning Zhang and Kai Xu and Jun Du and Yinan Li and Yuling Jiao and Xingyao Wu and Wu Liu and Xiliang Lu and Huikai Xu and Yirong Jin and Ruixia Wang and Haifeng Yu and S. P. Zhao},
      year={2024},
      eprint={2406.12195},
      archivePrefix={arXiv},
      primaryClass={quant-ph},
      url={https://arxiv.org/abs/2406.12195}, 
}

@misc{kremer2025optimizingnoncliffordcountunitarysynthesis,
      title={Optimizing the non-Clifford-count in unitary synthesis using Reinforcement Learning}, 
      author={David Kremer and Ali Javadi-Abhari and Priyanka Mukhopadhyay},
      year={2025},
      eprint={2509.21709},
      archivePrefix={arXiv},
      primaryClass={quant-ph},
      url={https://arxiv.org/abs/2509.21709}, 
}

@misc{wierichs_recursive_2025,
	title = {Recursive {Cartan} decompositions for unitary synthesis},
	url = {http://arxiv.org/abs/2503.19014},
	doi = {10.48550/arXiv.2503.19014},
	urldate = {2025-10-23},
	publisher = {arXiv},
	author = {Wierichs, David and West, Maxwell and Forestano, Roy T. and Cerezo, M. and Killoran, Nathan},
	month = jun,
	year = {2025},
	note = {arXiv:2503.19014},
}

@article{cartan_sur_1926,
	title = {Sur une classe remarquable d'espaces de {Riemann}},
	volume = {2},
	issn = {0037-9484, 2102-622X},
	url = {http://www.numdam.org/item?id=BSMF_1926__54__214_0},
	doi = {10.24033/bsmf.1105},
	urldate = {2025-10-23},
	journal = {Bulletin de la Société Mathématique de France},
	author = {Cartan, E.},
	year = {1926},
	pages = {214--264},
}

@article{drury_constructive_2008,
	title = {Constructive quantum {Shannon} decomposition from {Cartan} involutions},
	volume = {41},
	issn = {1751-8113, 1751-8121},
	url = {https://iopscience.iop.org/article/10.1088/1751-8113/41/39/395305},
	doi = {10.1088/1751-8113/41/39/395305},
	number = {39},
	urldate = {2025-10-23},
	journal = {Journal of Physics A: Mathematical and Theoretical},
	author = {Drury, Byron and Love, Peter},
	month = oct,
	year = {2008},
	pages = {395305},
}

@misc{BQSKit,
title = {Berkeley Quantum Synthesis Toolkit (BQSKit) v1},
author = {Younis, Ed and Iancu, Costin C. and Lavrijsen, Wim and Davis, Marc and Smith, Ethan},
abstractNote = {The Berkeley Quantum Synthesis Toolkit (BQSKit) is an optimizing quantum compiler and research vehicle that combines ideas from several projects at LBNL into one easily accessible and quickly extensible software package. The ideas in the QFAST, QSearch, LEAP, and QFactor software tools (all licensed through ipo.lbl.gov) all build upon one another. By combining these into one package, we create symbiotic interactions between the tools. This means better results, better throughput, less to maintain, and greater surface area to the public. Additionally, the BQSKit tool will create a research platform for future work here at LBNL.},
doi = {10.11578/dc.20210603.2},
url = {https://doi.org/10.11578/dc.20210603.2},
howpublished = {[Computer Software] \url{https://doi.org/10.11578/dc.20210603.2}},
year = {2021},
month = {apr}
}

@article{stable-baselines3,
  author  = {Antonin Raffin and Ashley Hill and Adam Gleave and Anssi Kanervisto and Maximilian Ernestus and Noah Dormann},
  title   = {Stable-Baselines3: Reliable Reinforcement Learning Implementations},
  journal = {Journal of Machine Learning Research},
  year    = {2021},
  volume  = {22},
  number  = {268},
  pages   = {1-8},
  url     = {http://jmlr.org/papers/v22/20-1364.html}
}

@misc{qiskit,
	title = {Quantum computing with {Qiskit}},
	url = {http://arxiv.org/abs/2405.08810},
	doi = {10.48550/arXiv.2405.08810},
	urldate = {2025-10-24},
	publisher = {arXiv},
	author = {Javadi-Abhari, Ali and Treinish, Matthew and Krsulich, Kevin and Wood, Christopher J. and Lishman, Jake and Gacon, Julien and Martiel, Simon and Nation, Paul D. and Bishop, Lev S. and Cross, Andrew W. and Johnson, Blake R. and Gambetta, Jay M.},
	month = jun,
	year = {2024},
	note = {arXiv:2405.08810},
}

@misc{PPO,
	title = {Proximal {Policy} {Optimization} {Algorithms}},
	url = {http://arxiv.org/abs/1707.06347},
	doi = {10.48550/arXiv.1707.06347},
	urldate = {2025-10-24},
	publisher = {arXiv},
	author = {Schulman, John and Wolski, Filip and Dhariwal, Prafulla and Radford, Alec and Klimov, Oleg},
	month = aug,
	year = {2017},
	note = {arXiv:1707.06347},
}

@article{magic_state_t,
	title = {Efficient magic state factories with a catalyzed {\textbar} {C} {C} {Z} ⟩ to 2 {\textbar} {T} ⟩ transformation},
	volume = {3},
	copyright = {https://creativecommons.org/licenses/by/4.0/},
	issn = {2521-327X},
	url = {https://quantum-journal.org/papers/q-2019-04-30-135/},
	doi = {10.22331/q-2019-04-30-135},
	language = {en},
	urldate = {2025-10-27},
	journal = {Quantum},
	author = {Gidney, Craig and Fowler, Austin G.},
	month = apr,
	year = {2019},
	pages = {135},
}

@article{moro_quantum_2021,
	title = {Quantum compiling by deep reinforcement learning},
	volume = {4},
	issn = {2399-3650},
	url = {https://www.nature.com/articles/s42005-021-00684-3},
	doi = {10.1038/s42005-021-00684-3},
	language = {en},
	number = {1},
	urldate = {2025-10-27},
	journal = {Communications Physics},
	author = {Moro, Lorenzo and Paris, Matteo G. A. and Restelli, Marcello and Prati, Enrico},
	month = aug,
	year = {2021},
	pages = {178},
}

@inproceedings{quetschlich_compiler_2023,
	address = {San Francisco, CA, USA},
	title = {Compiler {Optimization} for {Quantum} {Computing} {Using} {Reinforcement} {Learning}},
	copyright = {https://doi.org/10.15223/policy-029},
	isbn = {9798350323481},
	url = {https://ieeexplore.ieee.org/document/10248002/},
	doi = {10.1109/DAC56929.2023.10248002},
	urldate = {2025-10-27},
	booktitle = {2023 60th {ACM}/{IEEE} {Design} {Automation} {Conference} ({DAC})},
	publisher = {IEEE},
	author = {Quetschlich, Nils and Burgholzer, Lukas and Wille, Robert},
	month = jul,
	year = {2023},
	pages = {1--6},
}

@article{rosenhahn_monte_2023,
	title = {Monte {Carlo} graph search for quantum circuit optimization},
	volume = {108},
	issn = {2469-9926, 2469-9934},
	url = {https://link.aps.org/doi/10.1103/PhysRevA.108.062615},
	doi = {10.1103/PhysRevA.108.062615},
	language = {en},
	number = {6},
	urldate = {2025-10-27},
	journal = {Physical Review A},
	author = {Rosenhahn, Bodo and Osborne, Tobias J.},
	month = dec,
	year = {2023},
	pages = {062615},
}

@article{bilkis2021semi,
  title={A semi-agnostic ansatz with variable structure for variational quantum algorithms},
  author={Bilkis, Matias and Cerezo, M and Verdon, Guillaume and Coles, Patrick J and Cincio, Lukasz},
  journal={Quantum Machine Intelligence},
  volume={5},
  number={2},
  pages={43},
  year={2023},
  publisher={Springer},
url={https://link.springer.com/article/10.1007/s42484-023-00132-1},
doi={10.1007/s42484-023-00132-1}
}

@article{casas2025matchgate,
  title={Matchgate synthesis via Clifford matchgates and $T$ gates},
  author={ Casas, Berta and Braccia, Paolo and Gouzien, \'Elie and  Cerezo, M. and Garc\'ia-Mart\'in,, Diego},
  journal={Manuscript in preparation},
year={2025}
}

@misc{lizzio2025quantum,
author={Daniele Lizzio Bosco},
title = {Quantum Pre-Synthesis},
url = {https://github.com/Dan-LB/Quantum-Pre-Synthesis},
year = {2025}
}

@article{gibbs2025learning,
  title={Learning Circuits with Infinite Tensor Networks},
  author={Gibbs, Joe and Cincio, Lukasz},
  journal={arXiv preprint arXiv:2506.02105},
  year={2025},
url={https://arxiv.org/abs/2506.02105},
doi={10.48550/arXiv.2506.02105}
}

@article{caro2022outofdistribution,
  title={Out-of-distribution generalization for learning quantum dynamics},
  author={Caro, Matthias C and Huang, Hsin-Yuan and Ezzell, Nicholas and Gibbs, Joe and Sornborger, Andrew T and Cincio, Lukasz and Coles, Patrick J and Holmes, Zo{\"e}},
  journal={Nature Communications},
  volume={14},
  number={1},
  pages={3751},
  year={2023},
  publisher={Nature Publishing Group UK London},
url={https://www.nature.com/articles/s41467-023-39381-w},
doi={10.1038/s41467-023-39381-w}
}

@article{caro2021generalization,
  title={Generalization in quantum machine learning from few training data},
  author={Caro, Matthias C and Huang, Hsin-Yuan and Cerezo, M and Sharma, Kunal and Sornborger, Andrew and Cincio, Lukasz and Coles, Patrick J},
  journal={Nature {C}ommunications},
  volume={13},
  eid={4919},
  year={2022},
  publisher={Nature Publishing Group},
  doi={10.1038/s41467-022-32550-3},
url={https://www.nature.com/articles/s41467-022-32550-3}
}

@article{vidal2007classical,
  title={Classical simulation of infinite-size quantum lattice systems in one spatial dimension},
  author={Vidal, Guifr{\'e}},
  journal={Physical Review Letters},
  volume={98},
  number={7},
  pages={070201},
  year={2007},
  publisher={APS},
  doi={10.1103/PhysRevLett.98.070201},
url={https://journals.aps.org/prl/abstract/10.1103/PhysRevLett.98.070201}
}

@article{campbell2017roads,
  title={Roads towards fault-tolerant universal quantum computation},
  author={Campbell, Earl T and Terhal, Barbara M and Vuillot, Christophe},
  journal={Nature},
  volume={549},
  number={7671},
  pages={172--179},
  year={2017},
  publisher={Nature Publishing Group UK London},
  doi={https://doi.org/10.1038/nature23460},
  url={https://www.nature.com/articles/nature23460}
}

@article{horsman2012surface,
  title={Surface code quantum computing by lattice surgery},
  author={Horsman, Clare and Fowler, Austin G and Devitt, Simon and Van Meter, Rodney},
  journal={New Journal of Physics},
  volume={14},
  number={12},
  pages={123011},
  year={2012},
  publisher={IOP Publishing},
  doi={10.1088/1367-2630/14/12/123011},
url={https://iopscience.iop.org/article/10.1088/1367-2630/14/12/123011}
}

@article{bravyi2005universal,
  title = {Universal quantum computation with ideal Clifford gates and noisy ancillas},
  author = {Bravyi, Sergey and Kitaev, Alexei},
  journal = {Phys. Rev. A},
  volume = {71},
  issue = {2},
  pages = {022316},
  numpages = {14},
  year = {2005},
  month = {Feb},
  publisher = {American Physical Society},
  doi = {10.1103/PhysRevA.71.022316},
  url = {https://link.aps.org/doi/10.1103/PhysRevA.71.022316}
}

@inproceedings{RL_x_VQC_21,
author = {Ostaszewski, Mateusz and Trenkwalder, Lea M. and Masarczyk, Wojciech and Scerri, Eleanor and Dunjko, Vedran},
title = {Reinforcement learning for optimization of variational quantum circuit architectures},
year = {2021},
isbn = {9781713845393},
publisher = {Curran Associates Inc.},
address = {Red Hook, NY, USA},
booktitle = {Proceedings of the 35th International Conference on Neural Information Processing Systems},
articleno = {1391},
numpages = {13},
series = {NIPS '21},
url={https://proceedings.neurips.cc/paper_files/paper/2021/hash/9724412729185d53a2e3e7f889d9f057-Abstract.html}
}

@article{quantum_synthesis_quer_2021,
  author       = {Gregory Rosenthal},
  title        = {Query and Depth Upper Bounds for Quantum Unitaries via Grover Search},
  journal      = {CoRR},
  volume       = {abs/2111.07992},
  year         = {2021},
  url          = {https://arxiv.org/abs/2111.07992},
  eprinttype    = {arXiv},
  eprint       = {2111.07992},
  bibsource    = {dblp computer science bibliography, https://dblp.org}
}

@article{sun_asymptotically_2023,
	title = {Asymptotically {Optimal} {Circuit} {Depth} for {Quantum} {State} {Preparation} and {General} {Unitary} {Synthesis}},
	volume = {42},
	copyright = {https://ieeexplore.ieee.org/Xplorehelp/downloads/license-information/IEEE.html},
	issn = {0278-0070, 1937-4151},
	url = {https://ieeexplore.ieee.org/document/10044235/},
	doi = {10.1109/TCAD.2023.3244885},
	number = {10},
	urldate = {2025-10-27},
	journal = {IEEE Transactions on Computer-Aided Design of Integrated Circuits and Systems},
	author = {Sun, Xiaoming and Tian, Guojing and Yang, Shuai and Yuan, Pei and Zhang, Shengyu},
	month = oct,
	year = {2023},
	pages = {3301--3314},
}

@article{matchgate_circuit,
  title={Matchgates and classical simulation of quantum circuits},
  author={Richard Jozsa and Akimasa Miyake},
  journal={Proceedings of the Royal Society A: Mathematical, Physical and Engineering Sciences},
  year={2008},
  volume={464},
  pages={3089 - 3106},
  url={https://api.semanticscholar.org/CorpusID:10967491}
}

@ARTICLE{intro_RL,
  author={Sutton, R.S. and Barto, A.G.},
  journal={IEEE Transactions on Neural Networks}, 
  title={Reinforcement Learning: An Introduction}, 
  year={1998},
  volume={9},
  number={5},
  pages={1054-1054},
  doi={10.1109/TNN.1998.712192}}

@misc{braccia_optimal_2025,
	title = {Optimal {Haar} random fermionic linear optics circuits},
	url = {http://arxiv.org/abs/2505.24212},
	doi = {10.48550/arXiv.2505.24212},
	urldate = {2025-12-11},
	publisher = {arXiv},
	author = {Braccia, Paolo and Diaz, N. L. and Larocca, Martin and Cerezo, M. and García-Martín, Diego},
	month = may,
	year = {2025},
	note = {arXiv:2505.24212},
}

@misc{silver_mastering_2017,
	title = {Mastering {Chess} and {Shogi} by {Self}-{Play} with a {General} {Reinforcement} {Learning} {Algorithm}},
	url = {http://arxiv.org/abs/1712.01815},
	doi = {10.48550/arXiv.1712.01815},
	urldate = {2025-12-11},
	publisher = {arXiv},
	author = {Silver, David and Hubert, Thomas and Schrittwieser, Julian and Antonoglou, Ioannis and Lai, Matthew and Guez, Arthur and Lanctot, Marc and Sifre, Laurent and Kumaran, Dharshan and Graepel, Thore and Lillicrap, Timothy and Simonyan, Karen and Hassabis, Demis},
	month = dec,
	year = {2017},
	note = {arXiv:1712.01815},
}

@misc{schrittwieser_mastering_2020,
	title = {Mastering {Atari}, {Go}, {Chess} and {Shogi} by {Planning} with a {Learned} {Model}},
	url = {http://arxiv.org/abs/1911.08265},
	doi = {10.48550/arXiv.1911.08265},
	urldate = {2025-12-11},
	publisher = {arXiv},
	author = {Schrittwieser, Julian and Antonoglou, Ioannis and Hubert, Thomas and Simonyan, Karen and Sifre, Laurent and Schmitt, Simon and Guez, Arthur and Lockhart, Edward and Hassabis, Demis and Graepel, Thore and Lillicrap, Timothy and Silver, David},
	month = feb,
	year = {2020},
	note = {arXiv:1911.08265},
}

@misc{narvekar2020curriculumlearningreinforcementlearning,
      title={Curriculum Learning for Reinforcement Learning Domains: A Framework and Survey}, 
      author={Sanmit Narvekar and Bei Peng and Matteo Leonetti and Jivko Sinapov and Matthew E. Taylor and Peter Stone},
      year={2020},
      eprint={2003.04960},
      archivePrefix={arXiv},
      primaryClass={cs.LG},
      url={https://arxiv.org/abs/2003.04960}, 
}

\end{document}